\documentclass[aps,pra,epsf,superscriptaddress,amsmath,amssymb,amsfonts,twocolumn,showpacs,nofootinbib]{revtex4-1}

\usepackage{graphicx}
\usepackage{epsfig}
\usepackage{dcolumn}
\usepackage{bm}
\usepackage{braket}
\usepackage{amsmath}
\usepackage{mathtools}
\usepackage{graphicx,color,xcolor}
\usepackage{hyperref}
\newcommand{\abs}[1]{\left| #1 \right|} 
\usepackage{hyperref}
\usepackage{multirow}
\DeclareMathOperator{\sech}{sech} 

\usepackage[normalem]{ulem} 

\begin{document}

\title{Particle imbalanced weakly interacting  quantum droplets in one-dimension}

\author{I. A. Englezos}
\affiliation{Center for Optical Quantum Technologies, Department of Physics, University of Hamburg, 
Luruper Chaussee 149, 22761 Hamburg Germany}
\author{P. Schmelcher}
\affiliation{Center for Optical Quantum Technologies, Department of Physics, University of Hamburg, 
Luruper Chaussee 149, 22761 Hamburg Germany} \affiliation{The Hamburg Centre for Ultrafast Imaging,
University of Hamburg, Luruper Chaussee 149, 22761 Hamburg, Germany} 
\author{S. I. Mistakidis}
\affiliation{ITAMP, Center for Astrophysics $|$ Harvard $\&$ Smithsonian, Cambridge, MA 02138 USA} 
\affiliation{Department of Physics, Harvard University, Cambridge, Massachusetts 02138, USA}
\affiliation{Department of Physics, Missouri University of Science and Technology, Rolla, MO 65409, USA}

\date{\today}

\begin{abstract} 

We explore the formation of one-dimensional two-component quantum droplets with intercomponent particle imbalance using an \textit{ab-initio} many-body method. 
It is shown that for moderate particle imbalance each component maintains its droplet flat-top or Gaussian type character depending on the intercomponent attraction. 
Importantly, large particle imbalance leads to a flat-top shape of the majority component with the minority exhibiting spatially localized configurations.  
The latter imprint modulations on the majority component which become more pronounced for  increasing interspecies attraction. 
The same holds for larger mass or increasing repulsion of the minority species. 
Such structural transitions are also evident in the underlying two-body correlation functions.
To interpret the origin and characteristics of these droplet states we derive an effective model based on the established Lee-Huang-Yang theory providing adequate qualitative analytical predictions even away from its expected parametric region of validity. 
In contrast, the droplet character is found to vanish in the presence of fermionic minority atoms. 
Our results pave the way for unveiling complex droplet phases of matter.

\end{abstract}

\maketitle

\section{Introduction} 

Correlated quantum many-body states can be nowdays designed and experimentally prepared within ultracold atom platforms~\cite{BlochNature2012}. 
Prototypical examples are self-bound quantum droplets~\cite{Petrov2015,KadauDropExp,PfauReview,MalomedLuoReview,MalomedReview} and quasiparticles such as polarons~\cite{PolaronsBruun,PolaronsDemler}. 
The experimental observation of Bose~\cite{BosePolaronExp1,BosePolaronExp2,BosePolaronExp3,BosePolaronExp4,BosePolaronExp5} and Fermi polarons~\cite{FermiPolaronExp1,FermiPolaronExp2,FermiPolaronExp3} verified the crucial role of correlations in these settings. 
In turn, significant theoretical attention~\cite{BosePolaronDemler,BosePolaronFieldTheorySchidt} has been devoted towards the study of correlation effects in the stationary~\cite{PolaronsJaksch,PolaronsZinner,PolaronsGiorgini,PolaronsDemler2017,PolaronsTempere} and the far less explored non-equilibrium dynamics~\cite{PolaronDynamics1,PolaronDynamics2,PolaronDynamics3,PolaronDynamics4} of polarons. 

Higher-order correlations are similarly integral in the formation of quantum droplets. 
The latter can appear in short-range interacting bosonic mixtures~\cite{CabreraTarruellDropExp,CheineyTarruellDropExp,FortHeteroExp} but also in single-component~\cite{KadauDropExp,Bottcher2019SupersolidDrop,Chomaz_2023} and mixtures~\cite{Bisset2021,Smith2021} of dipolar gases. 
Droplet states manifest when quantum fluctuations, commonly accounted by the Lee-Huang-Yang (LHY) correction term~\cite{LeeHuangYang1957}, stabilize the gas against collapse originating from mean-field interaction effects~\cite{Petrov2015,PfauReview,MalomedLuoReview,MalomedReview}. 
Interestingly, other theory proposals for droplets suggest their occurrence in the presence of three-body interactions~\cite{Nishida3body,Morera3body1D} but also in Bose-Fermi mixtures with~\cite{CuiSpinOrbitBoseFermi} and without spin-orbit coupling~\cite{GajdaBoseFermi,Wang_2020BoseFermi}. 
Quantum droplets exhibit features of a dilute liquid like state~\cite{Petrov2015}, manifesting, for example, in the development of a flat-top (FT) profile in their spatial density configuration.  
In three dimensions (3D), the quantum liquid character of droplets can be unveiled in terms of their surface tension~\cite{Petrov2015,AncilottoLocalDensity} and incompressibility~\cite{FattoriCollisions}. 
However, the study of 3D droplets is hindered by their characteristic self-evaporation process~\cite{Petrov2015,FortModugnoSelfEvaporation}.
The later is absent in one-dimensional (1D) systems~\cite{AstrakharchikMalomed1DDynamics}, which emerge as ideal settings for studying long lived (due to lower densities) and stable quantum droplets, where correlation effects are naturally enhanced.

The majority of the quantum droplet investigations focused on imposing the fixed density ratio condition between the two components determined by their intracomponent interaction strengths.
In this regime, droplets are expected to be more stable  and in fact the two-component setting  is reduced to an effective single-component one~\cite{Petrov2015,SemeghiniFattoriDropExp}. 
Recently, some attention has been placed on exploring the genuine two-component nature of the system by employing either mass imbalance or different intracomponent interactions~\cite{Mistakidis2021,Englezos2023,FortHeteroExp}.
These studies indicated that quantum droplets can also  remain stable when the fixed density ratio is violated.
This has been further explored in 3D~\cite{QuantumCrit2023,FlynnPRR2023,FlynnTrapped2023} within the context of the LHY theory. 
Similar conclusions were drawn in 1D systems of strongly interacting lattice trapped particle imbalanced bosonic mixtures using the density matrix renormalization group method~\cite{latticeDrop2023} and in a ring geometry focusing on the rotational properties of droplets within LHY theory~\cite{tengstrand2022droplet}. 
Therefore, the stability region of droplets has been  extended to a wider range of density ratios relying on small intercomponent particle imbalances. 
However, considering larger particle imbalances shares the premise of constructing effective methods, through which analytical predictions can be made, but also enhance interparticle correlations (beyond LHY) since one of the components can be even reduced to a few-body sample. 
Note, also, that the works mentioned above~\cite{FlynnPRR2023,FlynnTrapped2023,latticeDrop2023,tengstrand2022droplet} explored the scenario of mass balanced mixtures featuring equal intraspecies interactions, leaving the number of particles per component as the only source of intercomponent imbalance. 

For these reasons, we focus on two-component short-range weakly interacting bosonic mixtures in 1D aiming to understand the interplay of particle, intracomponent interaction and mass imbalance on the droplet formation. 
In this context, one component (majority) consists of a significant larger number of atoms than the second (minority). 
We derive an effective model based on the coupled system of extended-Gross-Pitaevskii equations (eGPEs)~\cite{Petrov2015,PetrovLowD} which allows for analytical insights into the two-component droplet formation. 
For instance, it predicts a decoupling of the two-component system into a quantum droplet for the majority component and a localized bright-soliton structure for the minority one in the case of extreme atom imbalance.  
The results obtained from this effective theory are verified utilizing the {\it ab-initio} multi-layer multi-configuration time-dependent Hartree method for atomic mixtures (ML-MCTDHX)~\cite{Kronke_2013,Cao2013,cao2017unified,mistakidis2023few,lode2020colloquium}. 
Furthermore, the impact of increasing the involved interaction strengths or the mass of the minority atoms is revealed. It predominantly results in spatial modulations of the majority component which maintains a FT droplet profile. 
Next, it is shown that the droplet character of the majority bosonic subsystem vanishes in the presence of fermionic minority atoms, which in turn delocalize and spread over the former. 
Our results illustrate the surprising effectiveness of the LHY-theory in capturing droplet formation even far from its expected validity region.

This work is structured in the following way. In Section~\ref{sec:setup}, we describe the two-component attractively interacting bosonic mixture supporting droplet solutions. Section~\ref{sec:approaches} briefly introduces the underlying LHY theory as well as the nonperturbative ML-MCTDHX approach deployed for the investigation of quantum droplets.
Section~\ref{sec:eGPEResults} is devoted to the derivation and solution of the effective eGPEs in the limit of large intercomponent particle imbalance. It is used, later on, as an interpretation tool for the two-component droplet configurations.  
The phenomenology provided by the eGPEs is confirmed by comparing to the predictions of the {\it ab-initio} ML-MCTDHX method in both the one-[Sec.~\ref{sec:MBGSResults}] and two-[Sec.~\ref{sec:2bground}] particle level.
Droplet structures appearing in heteronuclear (Bose-Bose or Bose-Fermi) mixtures are briefly addressed in Section~\ref{sec:hetero}.  
Conclusions and possible future research directions are offered in Section~\ref{sec:SummaryAndOutlook}.

\section{Particle Imbalanced bosonic mixture}\label{sec:setup}

We employ a highly particle imbalanced bosonic mixture containing $N_A<N_B$ atoms of mass $m_{\sigma}$ ($\sigma=A,B$) and being confined in a weak 1D harmonic trap. 
Such a setting can be readily prepared via the technique of radiofrequency spectroscopy utilizing a two-photon Raman transition, where a portion of the atoms initial prepared in a single hyperfine state (e.g. $\ket{F=1,m_F=-1}$ of $^{39}$K) are transferred to another hyperfine state (e.g. $\ket{F=1,m_F=0}$ of $^{39}$K). 
Consequently, the percentage of atoms in each state can be controlled through the amplitude and Rabi-frequency of the applied  pulse~\cite{CabreraTarruellDropExp,CheineyTarruellDropExp,SemeghiniFattoriDropExp,BakkaliTowns2021}. 
The mixture is at cold temperatures where $s$-wave scattering is the predominant scattering process~\cite{olshanii1998atomic} and thus interactions are modelled by contact potentials. 
The inter-particle interactions are characterized by effective repulsive intra- ($g_{A}>0$, $g_{B}>0$) and attractive inter-component ($g_{AB}<0$) coupling strengths. 
These coefficients can be experimentally tuned through Feshbach resonances~\cite{chin2010feshbach,kohler2006production} via an external homogeneous magnetic field or confinement induced resonances~\cite{olshanii1998atomic} by means of manipulating the transversal trapping frequency. 

The resulting many-body Hamiltonian has the form 
\begin{align} \label{MB_Hamilt}
\begin{split}
&H = \sum_{\sigma=A,B} \sum_{i=1}^{N_{\sigma}} \left(-\frac{\hslash^2}{2m_\sigma}   \left(\frac{\partial^2}{\partial {x_i^\sigma}^2}\right)   + \frac{1}{2}m\omega^2(x_i^\sigma)^2\right) \\
&+ \sum_{\sigma=A,B}g_{\sigma} \sum_{i<j}^{N_{\sigma}} \delta(x_i^{\sigma} -x_j^{\sigma}) + g_{AB} \sum_{i=1}^{N_{A}} \sum_{j=1}^{N_B} \delta(x_i^A -x_j^B).    
\end{split}
\end{align}
In order to ensure the 1D nature of the ensuing dynamics we consider a fixed and adequately large aspect ratio between the longitudinal ($\omega_{{\rm x}}$) and the transverse ($\omega_{{\rm \perp}}$) trapping frequencies. 
Specifically, $\omega=\omega_{{\rm x}}/\omega_{{\rm \perp}}=0.01$ which is typical in 1D experiments~\cite{Ketterle2001LowDexp,romero2023experimental} and prevents the involvement of transversal excitations. 
Finally, for computational convenience we rescale the above Hamiltonian with respect to $\hslash\omega_\perp $.
Hence, the length, time and interaction strengths are given in units of $a_{\perp}=\sqrt{\hslash/(m\omega_{\rm \perp}) }$, $1/\omega_{\rm \perp}$ and $\sqrt{\hslash^3\omega_{\rm \perp} /m}$ respectively.

\section{Many-body description}\label{sec:approaches} 

\subsection{Extended Gross-Pitaevskii equations} 

Two-component, homonuclear $(m_A=m_B \equiv m)$, 1D quantum droplets in the presence of the first-order quantum correction (LHY contribution) are described, in the weakly interacting regime, by the following coupled eGPEs~\cite{PetrovLowD,MithunMI}: 

\begin{subequations}\label{eGPE}
\begin{align}
i\hslash \frac{\partial \Psi_A(x,t)}{\partial t}=&\Bigg [  - \frac{\hslash^2}{2m}\frac{\partial^2 }{\partial x^2} + G_A |\Psi_A|^2 \notag\\
&-(1 -G)g |\Psi_B|^2 + V(x) \label{eq: eGPEa}\\ 
&- \frac{g_A\sqrt{m}}{\pi\hslash} \sqrt{g_A|\Psi_A|^2 +g_B|\Psi_B|^2 } \Bigg ] \Psi_A(x,t), \notag\\ 
&\notag\\
i\hslash \frac{\partial \Psi_B(x,t)}{\partial t}=&\Bigg [ - \frac{\hslash^2}{2m}\frac{\partial^2 }{\partial x^2} + G_B |\Psi_B|^2 \notag \\
&-(1 -G)g |\Psi_A|^2 +V(x)  \label{eq: eGPEb}\\
&- \frac{g_B\sqrt{m}}{\pi\hslash} \sqrt{g_A|\Psi_A|^2 +g_B|\Psi_B|^2 }\Bigg ] \Psi_B(x,t). \notag
\end{align}
\end{subequations}
In these expressions, $G_A=g_A+Gg_B$, $G_B=g_B+Gg_A$, and  $G=2g\delta g/(g_A+g_B)^2$. 
The average mean-field repulsion is  $g=\sqrt{g_Ag_B}$, and the distance from the mean-field balance point is given by $\delta g = g + g_{\rm AB}$. 
Also, $V(x)$ represents the external trapping potential which we consider herein to be a harmonic trap as in the many-body Hamiltonian of Eq.~(\ref{MB_Hamilt}). 
For our simulations, we use the normalization condition $\int |\Psi_\sigma|^2 dx = N_\sigma$, where $\Psi_{\sigma}(x,t)$ represents the 1D wave function of the $\sigma=A, B$ component and $n_{\sigma}=\abs{\Psi_{\sigma}}^2$ is the respective density normalized to the atom number. 
Notice that for a balanced mixture, i.e. $|\Psi_A|^2\sqrt{g_B}=|\Psi_B|^2\sqrt{g_A}$, it is known that the two-component system is reduced to an effective single-component one~\cite{PetrovLowD}, where both components behave identically.  
In this case, a transition of the droplet density from a Gaussian to a FT configuration takes place for either increasing particle number ($N=N_A+N_B$) or decreasing intercomponent attraction ($|g_{\rm AB}|$) with $0< - g_{\rm AB} < g$~\cite{PetrovLowD}. 
Interestingly, due to the droplet incompressibility, the emergent FT structures exhibit a saturation peak density at $n_0 = 8g/(9\pi^2\delta g^2)$~\cite{PetrovLowD}.
However, it is not \textit{a-priori} expected that these droplet properties are retained in the particle-imbalanced two-component setting. This is one of the questions that we address below.   

It is also important at this point to explicate  the validity of the eGPEs~\eqref{eGPE}. This framework in principle holds for macroscopic systems, in free-space and close to the mean-field  balance point $\delta g \approx 0$~\cite{PetrovLowD}.
However, several works relying on non-perturbative methods have demonstrated that the eGPEs can provide accurate predictions, at least on the  qualitative level, even for mesoscopic systems and for non-vanishing $\delta g$~\cite{ParisiMonteCarlo2019,ParisiGiorginiMonteCarlo,Mistakidis2021} but also in the presence of a shallow ($\omega_x \ll \omega_{\perp}$) external trap~\cite{Englezos2023}.
Therefore, the eGPE framework has been shown to provide an accurate phenomenological description of quantum droplets, even for certain systems which it is not designed for. 
Throughout this work we study the crossover from particle balanced droplet configurations to strong intercomponent imbalance where the minority component tends to the impurity limit. 
As such, we explore parametric regimes lying far outside the expected  validity region of the eGPEs framework.
We showcase, however, by comparing with {\it ab-initio} calculations, that the eGPEs provides surprisingly valuable insights on the rich phenomenology exhibited close to the impurity limit ($N_A \ll N_B$).  

\begin{figure*}[ht]
\centering
\includegraphics[width=0.98\textwidth]{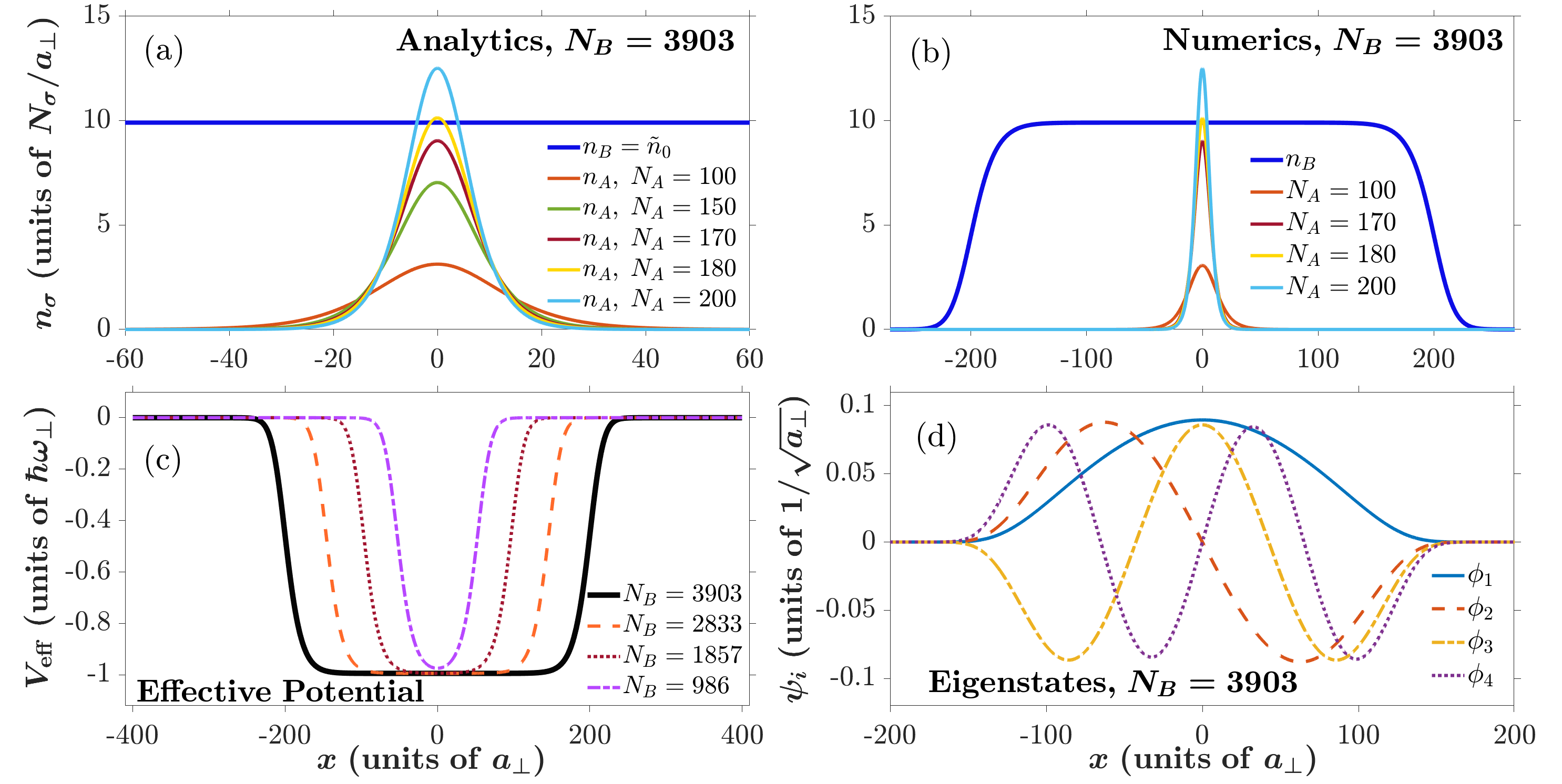}
\caption{Density profiles for fixed $N_B=3903$, $g_B=0.05$, $g_A=0.049$ and $g_{AB}=-0.149$ of (a) the analytical and (b) the numerical solutions to the reduced eGPEs~(\ref{ReducedeGPE}) for various particle numbers in the minority component (see legend). Upon increasing the particle number of the minority component $N_A$ its density maximum becomes comparable to the droplet saturation density, inspite of the large particle difference (see main text). In (c) we illustrate the effective potential experienced by the minority component for different atom numbers in the majority component (see legend). (d) The first four eigenstates of the effective potential of Eq.~(\ref{effPot}) for $N_B=3903$ (see legend) are depicted.}
\label{fig:GSmodel}
\end{figure*} 

\subsection{Many-body variational  approach}\label{sec:Variational_Method}

As discussed above, in order to judge the parametric regions of  validity of the eGPEs for mesoscopic mixtures but also identify the impact of beyond LHY correlations on the formation of particle-imbalanced droplets we independently rely on the {\it ab-initio} ML-MCTDHX method~\cite{Kronke_2013,Cao2013,cao2017unified}.  
Within this approach the full many-body wave function is expressed in a multi-layer structure. The latter  utilizes a variationally optimized time-dependent basis set in order to numerically solve the corresponding many-body Schr\"odinger equation~\cite{mistakidis2023few,lode2020colloquium}. 
Accordingly, the relevant Hilbert space is spanned efficiently and interparticle correlations are captured.

The intercomponent correlations (entanglement) of the bosonic mixture are taken into account through a truncated Schmidt decomposition~\cite{horodecki2009quantum}.
This way, $D$ different orthonormal species functions, $\ket{\Psi_k^\sigma(t)}$, are used for each component $\sigma=A,B$ and the many-body wave function reads    
\begin{equation}
\ket{\Psi(t)}= \sum_{k=1}^{D} \sqrt{\lambda_k(t)} \ket{\Psi_{k}^{A}(t)} \ket{\Psi_{k}^{B}(t)}.
\label{SpeciesFunctionLevel}
\end{equation} 
Here, the eigenvalues of the species reduced density matrix~\cite{mistakidis2018correlation,cao2017unified} are the time-dependent Schmidt weights $\sqrt{\lambda_k(t)}$ which determine the degree of intercomponent correlations. 
Namely, if at least two distinct $\lambda_k$'s are finite the many-body wave function is in a superposition and the system may be considered entangled~\cite{horodecki2009quantum,mistakidis2023few}.
However, in the case of $\lambda_1(t)=1$ and $\lambda_{k>1}(t)=0 $, the many-body ansatz is simply a product (non-entangled) state.

As a next step, intracomponent correlations are included by expanding each species function in terms of a linear superposition of time-dependent number states $\ket{\mathbf{n_k}^\sigma}$, 
\begin{align}
\ket{\Psi_{k}^{\sigma}(t)}= \sum_{\mathbf{n_k^\sigma}|N_k} A_{n_k}^{\sigma}(t) \ket{\mathbf{n}_k^\sigma}, 
\label{permenantStates}
\end{align}
with time-dependent expansion coefficients $A_{n_k}^{\sigma}(t)$.
These number states $\ket{\mathbf{n}_k^\sigma}$ correspond to the full set of permanents constructed by $d_\sigma$ time-dependent variationally optimized single-particle functions $\ket{\Phi_i^\sigma}$ with occupation numbers $\mathbf{n}=(n_1,...,n_{d_\sigma})$.
In turn, the $d_{\sigma}$ time-dependent single-particle functions evolve in the single-particle Hilbert space spanned by the time-independent (primitive) basis $\{ \ket{r_j^k} \}_{j=1}^{ \mathcal{M}}$.
In this study, the latter refers to a $\mathcal{M}$ dimensional discrete variable representation with $\mathcal{M}=1000$ grid points. 
Finally, the resulting equations of motion for the coefficients of the ML-MCTDHX wave function ansatz describing the many-body  Hamiltonian of Eq.~\eqref{MB_Hamilt} are found, for instance, by using the Dirac-Frenkel variational principle~\cite{frenkel1934wave,cao2017unified}, $\braket{\delta\Psi|( i\hbar \partial_t-\Hat{H})|\Psi}=0$.

Concluding, we note in passing that the ML-MCTDHX wave function ansatz easily reduces to the usual MF one that neglects all correlations~\cite{pethick2008bose}, $\ket{\Psi_{{\rm MF}}(t)}=\prod_{i=1}^{N_A}\ket{\Phi_i^A(t)} \prod_{i=1}^{N_B}\ket{\Phi_i^B(t)}$, by using $D=d_A=d_B=1$.  
Then the variational principle recovers the well-known coupled set of Gross-Pitaevskii equations for the bosonic mixture~\cite{pethick2008bose,Stringari2016BEC}.
However, the eGPEs~\eqref{eGPE} take into account correlations in a perturbative manner and thus do not follow directly from the ML-MCTDHX ansatz.
The latter incorporates beyond LHY correlations and therefore allows to determine whether the eGPE description is sufficient to capture the participating correlation effects, to a good approximation, or if higher-order ones become significant.  

\section{Effective description in the large Particle Imbalance limit}\label{sec:eGPEResults} 

We consider a two-component mixture in free space ($V(x)=0$), with attractive intercomponent interactions, featuring extreme particle imbalance between the two macroscopically occupied components, i.e. 
$N_B \gg N_A \gg 1$. 
In this case, we may keep in the eGPE description [Eq.~(\ref{eGPE})] only terms scaling at least as $\sqrt{N_B}$ ($\sqrt{N_A}$) in the majority (minority) component equation but ignore contributions  $\mathcal{O} (N_A/N_B)$ and $\mathcal{O} (N_A)$, respectively see also the discussion below. 
Then, assuming $\sqrt{N_B} \gg N_A$, the genuine two-component system of eGPEs~\eqref{eq: eGPEa}-\eqref{eq: eGPEb} reduces to
\begin{subequations}\label{ReducedeGPE}
\begin{align}
    i\hslash \frac{\partial \Psi_A}{\partial t}=&\Bigg [ - \frac{\hslash^2}{2m}\frac{\partial^2 }{\partial x^2} + V_{{\rm eff}}(|\Psi_B|) +G_A |\Psi_A|^2 \Bigg ]\Psi_A,\label{eq: a} \\
    &\notag\\
    i\hslash \frac{\partial \Psi_B}{\partial t}=&\Bigg [ - \frac{\hslash^2}{2m}\frac{\partial^2 }{\partial x^2} + G_B |\Psi_B|^2 - \frac{g_B\sqrt{m g_B}}{\pi\hslash} |\Psi_B| \Bigg ] \Psi_B. \label{eq: b}
\end{align}
\end{subequations}
Apparently, in this limit, the minority component experiences an effective potential created by the majority species of the form   
\begin{equation} \label{effPot}
V_{{\rm eff}}(|\Psi_B|)= - (1-G)g|\Psi_B|^2 - \frac{g_A\sqrt{mg_B}}{\pi\hslash}|\Psi_B|.
\end{equation}
It turns out that the majority component decouples from the minority and obeys a reduced single-component eGPE~\cite{PetrovLowD}. 
The latter contains modified effective non-linear interaction parameters determined by the presence of the minority component i.e. $\Tilde{\delta} g=G_B$ and $\Tilde{g}=g_B/2^{1/3}$. 
This indicates that the presence of the minority component induces a global effect on the majority one, i.e. extending beyond their overlap region. 
In this sense, the creation and structural configurations of droplet-like states in the majority component depend strongly on the characteristics of the minority one.  
According to the above, it is possible to control the saturation density and hence the overall behavior of the majority species, via tuning the interaction strengths associated with the minority component $(g_A, g_{AB})$, and therefore also $G_B$.

The majority component equation~\eqref{eq: b} has the well-known 1D droplet solution~\cite{PetrovLowD,MithunMI} (setting $\hslash =m=1$)
\begin{equation}
\label{DropletSolution}
\Psi_B(x,t) = \frac{\sqrt{\Tilde{n}_0 } ( \Tilde{\mu}/ \Tilde{\mu}_0) e^{-i\Tilde{\mu}t} }{ 1+ \sqrt{1- \Tilde{\mu}/ \Tilde{\mu}_0 } \cosh{\sqrt{-2m\Tilde{\mu}}x}},
\end{equation}
where $\Tilde{\mu}_0 = - \Tilde{\delta} g \Tilde{n}_0 /2 $ represents the minimum value of the chemical potential $\Tilde{\mu}$ 
above which droplet solutions exist~\cite{PetrovLowD,AstrakharchikMalomed1DDynamics}. 
For sufficiently large particle number, Eq.~\eqref{DropletSolution} predicts a FT density profile at the droplet saturation density $\Tilde{n}_0 = 8 \Tilde{g}^3/(9\pi^2\Tilde{\delta} g^2)$, see e.g. the thick blue lines in Fig.~\ref{fig:GSmodel}(a), (b).  
Using the droplet solution of Eq.~(\ref{DropletSolution}) we provide characteristic examples of the effective potential, $V_{\rm eff} (|\Psi_B|)$, experienced by the minority component for various atom numbers ($N_B$) in the majority component, see Fig.~\ref{fig:GSmodel}(c). 
As expected, $V_{\rm eff} (|\Psi_B|)$ exhibits an inverse droplet profile being reminiscent of a square-well with a pronounced flat potential minimum at large majority atom numbers, $N_B$, and transits towards a bell shaped inverted profile for decreasing $N_B$. 
The first four eigenstates of $V_{\rm eff} (|\Psi_B|)$, obtained numerically via diagonalization, when the droplet solution is deep in the FT regime ($N_B=3903$) are  provided in Fig.~\ref{fig:GSmodel}(d). 
Evidently, they are reminiscent of the eigenstates of a square well, featuring sinusoidal profiles, with a hierarchy in terms of their nodes for higher-lying ones, inside the FT region and rapidly decaying  at the tails of the droplet.

Since we consider $N_B \gg N_A \gg 1 $, i.e. operate close to the thermodynamic limit, it is natural to assume that the majority component is deep in the FT regime. 
Accordingly, its density profile acquires the constant saturation value $|\Psi_B(x)|^2 \approx \Tilde{n}_0$ away from the edges of the atomic cloud, and hence also in the comparatively much smaller spatial overlap region with the minority component. 
The latter then may be well approximated by a bright soliton solution of the form $\Psi_A (x) \approx A \sech{(\lambda x)}e^{-i\mu_A t}$ where $\mu_A$ is the chemical potential of the minority species. 
By substituting this solution into Eq.~\eqref{eq: a} we find 
\begin{equation}
    \label{bright}
    \begin{split}
        \lambda^2 =& -2m\Big[ (1-G)g \Tilde{n}_0 + \frac{g_A\sqrt{mg_B}}{\pi}\sqrt{\Tilde{n}_0} +\mu_A  \Big],\\
        |A|^2 = & \frac{-\lambda^2}{m(g_A+Gg_B)}.
    \end{split}
\end{equation}
This solution, $\Psi_A(x)$, determined through Eq.~\eqref{bright} predicts that the minority component becomes gradually more localized in space and features an increased amplitude for larger $N_A$ and fixed interaction coefficients as shown in Fig.~\ref{fig:GSmodel}(a). 
Hence, in spite of the pronounced intercomponent particle imbalance, there is a critical minority atom number at which $|A|^2\approx \Tilde{n}_0$. 
In this limit, the assumption of decoupled components ceases to be valid, see e.g. the light-blue or yellow lines depicted in Fig.~\ref{fig:GSmodel}(a), (b). 
Namely, using the normalization condition $\int |\Psi_A|^2 dx = N_A$ together with Eq.~\eqref{bright} we find the critical chemical potential $\mu_A^{\rm crit} = V_{\rm eff}(\sqrt{\Tilde{n}_0}) +G_A \Tilde{n}_0/2 $ and critical number of particles in the minority component $N_A^{\rm crit} = \sqrt{-\frac{4\Tilde{n}_0}{G_A m }}$, e.g. $N_A^{\rm crit} \approx 174$ for the parameter values used in Fig.~\ref{fig:GSmodel}. 
Beyond this point the majority component is expected to exhibit density modulations, on top of the FT, which are located at the overlap region with the minority component. 
To confirm the validity of the analytical soliton solution for the minority component in the case of a  highly particle imbalanced system we provide the ground state densities of the minority species, see Fig.~\ref{fig:GSmodel} (b), obtained from the simulation of the reduced single-component Eq.~\eqref{eq: a} assuming the FT solution of Eq.~\eqref{DropletSolution} for the majority component. A comparison of the wave forms depicted in Fig.~\ref{fig:GSmodel} (a) and (b) reveals an almost perfect agreement between the two approaches. 

\begin{figure*}[ht]
\centering
\includegraphics[width=0.98\textwidth]{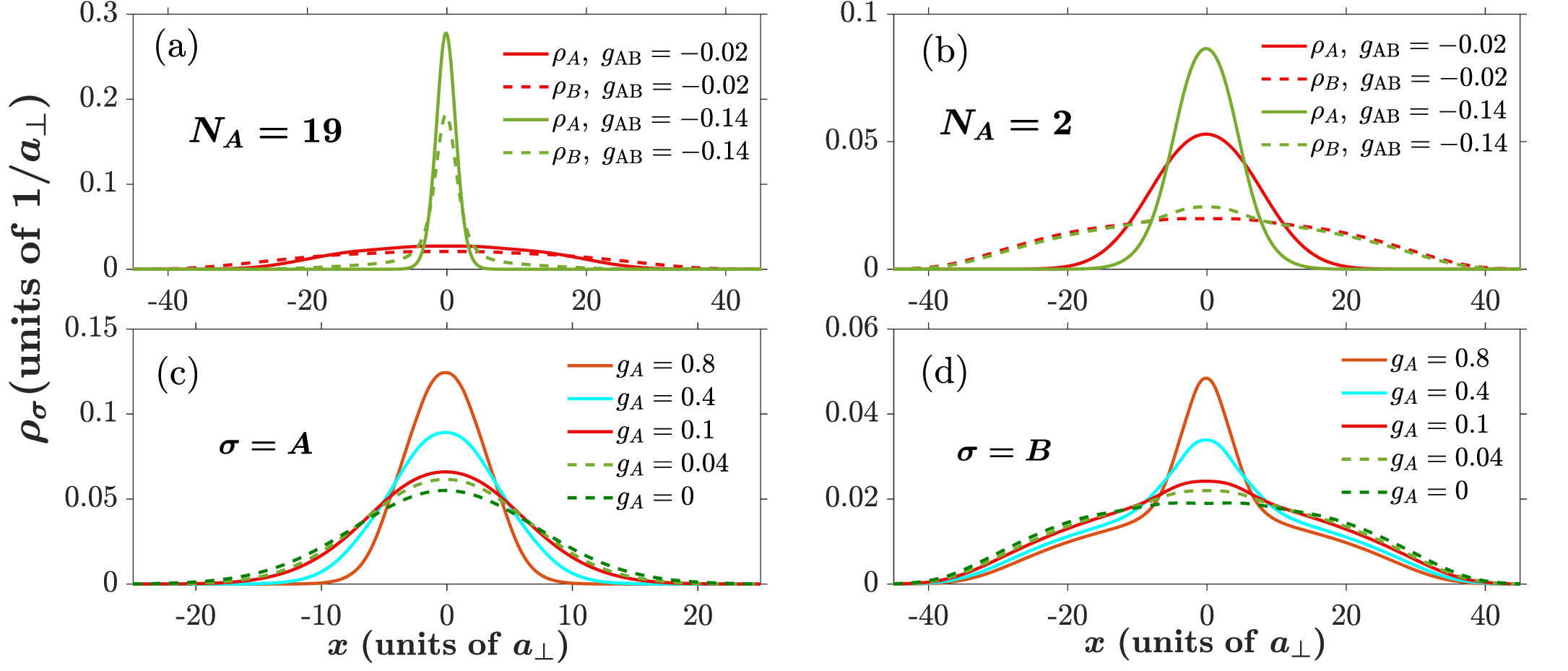}
\caption{Density profiles of the two-component bosonic droplet many-body ground state configurations for different intercomponent attractions $g_{AB}$ (see legends) and (a) $N_A = 19$, or (b) $N_A = 2$.  
(c) [(d)] Ground state densities of (c) [(d)] the minority [majority] component for $N_A=4$ and $\delta g =0.01$, while tuning the intracomponent repulsion of the minority species $g_A$ (see legends). 
For sufficiently small ratios $N_A/N_B$ the majority component maintains its droplet configuration, while modulations upon the FT profile arise with increasing intercomponent attraction or intracomponent repulsion of the minority species. In all cases, the remaining system parameters $N_B=40$ and $g_A=g_B=0.1$ are kept fixed (unless stated otherwise), while a shallow harmonic trap, $\omega= 0.01$, is used.}
\label{fig:GSImb}
\end{figure*}

To shed light on these density undulations of the majority component, we deploy the next-order correction to the one used for obtaining the reduced equations~\eqref{eq: a}-\eqref{eq: b}. It stems from the non-vanishing intercomponent particle number ratio $N_A/N_B\neq 0 $. 
This next-order correction, scaling as $N_A/\sqrt{N_B}$, originates from the LHY-term and it is given by $-g_{A [B]}\frac{g_A|\Psi_A|^2}{2\sqrt{g_B}|\Psi_B|}\Psi_{A [B]}$ for each component respectively. 
Evidently, this term manifests a direct coupling among the components and it is responsible for the aforementioned density modulations on the FT profile of the majority component. These modulations are predominantly enhanced for increasing either $N_A$ or $g_A$. 
The same holds for larger $g_B$ or smaller $N_B$ but in a 'slower' manner since the underlying scaling is of square root type. 

For completeness, we note that the next higher-order correction term (scaling as $N_A$) to the majority species equation~\eqref{eq: b}, stems from the direct coupling to the minority component (i.e. $-(1 -G)g |\Psi_A|^2\Psi_B$) and it apparently also directly depends on $g_A$ and $N_A$. 
This term also depends on the interspecies interaction $g_{AB}$, and increasing $g_{AB}$ for fixed $g_A$ indeed results in enhanced modulations of the density of the majority component. However, as we shall explicate below, the impact of $g_{AB}$ appears to be less prominent as compared to the effect following an increase of $g_{A}$. 
Hence, $g_A$ is the most important interaction parameter for probing deviations with respect to the intercomponent decoupled limit characterized by Eq.~\eqref{eq: a}-\eqref{eq: b}. 
Finally, it should be emphasized that the effective description of Eq.~\eqref{eq: a}-\eqref{eq: b} should not be considered as an exact quantitative model especially so for finite systems. 
It is, however, a rather qualitative model in the thermodynamic limit, which can aid to the interpretation of the droplet behavior in finite systems that we explore in this work.

\section{Many-Body Ground State}\label{sec:MBGSResults}

Naturally, the striking two-component droplet behavior predicted within the effective eGPE framework [Eqs.~(\ref{ReducedeGPE})], needs to be verified by explicit many-body calculations especially so away from the thermodynamic limit.
Moreover, it is worth mentioning that we operate in a regime, where the condition $n_A/n_B\approx\sqrt{g_B/g_A}$ is violated, and thus lays outside the commonly considered parameter region of symmetric  droplets~\cite{QuantumCrit2023,latticeDrop2023,FlynnPRR2023,FlynnTrapped2023}.  

For this reason, we employ the {\it ab-initio} ML-MCTDHX method~\cite{cao2017unified,Mistakidis2021,Englezos2023} which allows to quantify the many-body properties of the system. 
To render this setup numerically tractable with an {\it ab-initio} method, we constrain the size of the majority component to mesoscopic [here  $N_B=40$ presented in Fig.~\ref{fig:GSImb} and  $N_B=20$ (not shown for brevity)]. Also, a weak harmonic trap, which is an experimentally common situation~\cite{Ketterle2001LowDexp}, characterized by $\omega=0.01$ is applied. 
Both of these restrictions result in droplet configurations having comparatively smaller FT density signatures (from their free space counterparts) located around the trap center as it was demonstrated, for instance, in Refs.~\cite{Englezos2023,ParisiGiorginiMonteCarlo}.  
However, as we shall explicate below, a close inspection of the corresponding density profiles $\rho_{\sigma}(x)=  \langle \Psi|\hat{\Psi}_\sigma ^\dagger (x)  \hat{\Psi}_\sigma (x) |\Psi \rangle$ (normalized to unity), as well as the intracomponent two-body correlation functions (see Sec.~\ref{sec:2bground}), allows us to identify the droplet-like character (being either FT or Gaussian shaped) of the majority component in the resulting many-body configurations. 
Also, we remark that all configurations to be presented below possess an energy per particle that is below the first trapped state (i.e. $E_{{\rm MB}}/(N_A+N_B)<\hbar\omega/2$), thus further confirming the bound state character of the ensuing many-body state.

First, we assume small particle imbalance, i.e. $N_B=40$, $N_A=19$, and study the underlying ground state configurations for different intercomponent attractions $g_{AB}$, see Fig.~\ref{fig:GSImb}(a).  
It becomes evident that due to the increasing attraction the component densities deform from spatially extended FT droplet structures to highly localized, soliton-type, profiles, see e.g. the green solid and dashed lines in Fig.~\ref{fig:GSImb}(a). 
The aforementioned transition behavior is also known to occur in the case of particle balanced mixtures and will eventually lead the system to  collapse
\footnote{By collapse in 1D we refer here to the increasing spatial localization, until the width of the density in the elongated direction becomes comparable to the transverse length scale, $a_{\perp}$. 
In this latter regime the assumption of the 1D setting is invalidated~\cite{pethick2008bose}.} for sufficiently strong intercomponent attraction~\cite{PetrovLowD,Mistakidis2021,pethick2008bose}. 
Nevertheless, the densities of both components closely follow each other with the majority species showing presignatures of extended tails that become pronounced for larger particle number ratios as we showcase below. 
In 3D it has been recently shown~\cite{QuantumCrit2023,FlynnPRR2023,FlynnTrapped2023} that in the case of moderate particle imbalance, such as the one portrayed in Fig.~\ref{fig:GSImb}(a), the system is characterized by either a bound imbalanced droplet or droplet-gas coexistence, due to the particle emission (or self evaporation) of 3D   droplets~\cite{Petrov2015,SemeghiniFattoriDropExp}.
This instability mechanism, however, is absent in 1D~\cite{PetrovLowD,Collective1D} at least within the weakly interaction regime. Hence, we observe the majority component maintaining its droplet character for all particle imbalances and as we shall argue later on also in terms of their two-body correlation patterns [see Sec.~\ref{sec:2bground}].

In sharp contrast, we observe that upon further decreasing the atom number in the minority subsystem (e.g. $N_B=40$, $N_A=2$), the majority component largely retains its FT  droplet configuration which is distorted only within  the spatial overlap region of the components, as long as $g_{AB} \neq 0$, see   Fig.~\ref{fig:GSImb}(b). 
On the other hand, the density of the minority component exhibits a soliton-type structure, exhibiting increased spatial localization for larger attractions.  
This behavior reaffirms the predictions of the reduced system of the eGPEs [Eqs.~\eqref{eq: a}-\eqref{eq: b} in  Sec.~\ref{sec:eGPEResults}] indicating that sufficiently large particle imbalance, i.e.  $N_B/N_A\gg 1$, prevents or at least delays the 1D collapse taking place for increasing attraction. This is a quite interesting mechanism that should be also testified in the quasi-1D system and it is thus a fruitful perspective for future investigations. 
Additionally, the majority component retains its FT droplet configuration (for large particle numbers $N_B\gg 1$), while the minority species develops a solitonic profile as dictated by Eq.~\eqref{ReducedeGPE}.

As a next step, we focus on the above-discussed large particle imbalanced system and explore its dependence on the minority species repulsion, $g_A$. 
Notice that previous works considering particle imbalanced droplets mainly assumed $g_A=g_B$~\cite{FlynnPRR2023,FlynnTrapped2023,latticeDrop2023,tengstrand2022droplet}. 
Hence, the understanding of the interplay between different sources of intercomponent imbalance is far from complete. 
Here, we address the impact of interaction imbalance, while in Sec.~\ref{sec:hetero} we discuss the effect of intercomponent mass imbalance in particle imbalanced droplets. 

Recall that in Sec.~\ref{sec:eGPEResults} we argued that the correction to the majority species equation, originating from the LHY term is $-\sqrt{g_{B}}g_A\frac{|\Psi_A|^2}{2|\Psi_B|}\Psi_{B}$ and it enforces coupling among the components. 
This implies that the spatial undulations of the majority component in its overlap region with the minority one strongly depend on $g_A$. 
Figures~\ref{fig:GSImb}(c), (d) present the density configurations of both the minority and the majority components respectively for various $g_A$ values.  
The minority component has a Gaussian density configuration [Fig.~\ref{fig:GSImb}(c)] becoming gradually more localized and tending towards a solitonic structure for increasing repulsion~\cite{SymbioticSolitons,Abdullaev2008}.  
Simultaneously, the majority component shows a modulated FT droplet structure for all values of $g_A$ and features more prominent modulations in the vicinity of the minority species for larger $g_A$ [Fig.~\ref{fig:GSImb}(d)].
Surprisingly, this behavior for increasing $g_A$ is in qualitative agreement with the predictions of the reduced eGPE model in the limit of large particle imbalance, despite the fact that the present setting lies outside the validity region of the eGPEs\eqref{eq: a}-\eqref{eq: b}. 
Notice also that the density modulations of the majority component on top of its FT profile essentially vanish for $g_A=0$.  
In particular, for $g_A=0$ and $N_A=4$ (see the dashed-green lines in Fig.~\ref{fig:GSImb}(c), (d)),   the majority component features a clear FT density profile  with no visible modulation even within the overlap region. This is in accordance with the decoupled scenario described by Eq.~\eqref{ReducedeGPE} and presented in Fig.~\ref{fig:GSmodel}(a). 

Moreover, we observe that tuning the intraspecies interaction strength $g_A$ (for fixed $\delta g$) has a more significant impact on the density profile of the majority component, as compared to adjusting the interspecies attraction $g_{AB}$ (for fixed $g_A$). 
This becomes evident by the relatively enhanced density modulations of the majority shown in Fig.~\ref{fig:GSImb}(c), (d) when contrasted to the ones in Fig.~\ref{fig:GSImb}(b). 
It is also consistent with the conclusions of the effective model, since both the correction originating from the LHY term (i.e. $-\sqrt{g_{B}}g_A\frac{|\Psi_A|^2}{2|\Psi_B|}\Psi_{B}$) and the one stemming from the direct coupling to the minority species (i.e $-(1-G)g|\Psi_A|^2\Psi_B$) explicitly depend on $g_{A}$. They also vanish for $g_A=0$, while $g_{AB}$ enters explicitly only on the latter.
In particular, the pre-factor of the latter (i.e. $-(1-G)g$) is reduced by a factor of two upon tuning the interspecies interaction strength from $g_{AB}=-0.02$ to $g_{AB}=-0.14$ (for fixed $g_A$), while the former scales linearly with $g_A$. 

It is worth noting here, that the opposite limit of strong intra and inter-component interactions (such that $\delta g$ is relatively small) and moderate particle imbalance was recently studied in Ref.~\cite{latticeDrop2023} utilizing an optical lattice.
Under these conditions, the system was found to feature imbalanced droplets at the overlap region between the two components, while excess particles remained in a gas or a super Tonks-Girardeau gas phase~\cite{latticeDrop2023}. 
This appears to be already consistent with our effective model  analysis. 
Apparently, for strong interactions the coupling terms discussed above dominate (instead of providing perturbative corrections as in the limit of weak interactions considered here).  
Hence, we expect the droplet or gas character of the mixture to be primarily characterized by the behavior at the overlap region between the two components.

\begin{figure}[ht]
\centering
\includegraphics[width=0.47\textwidth]{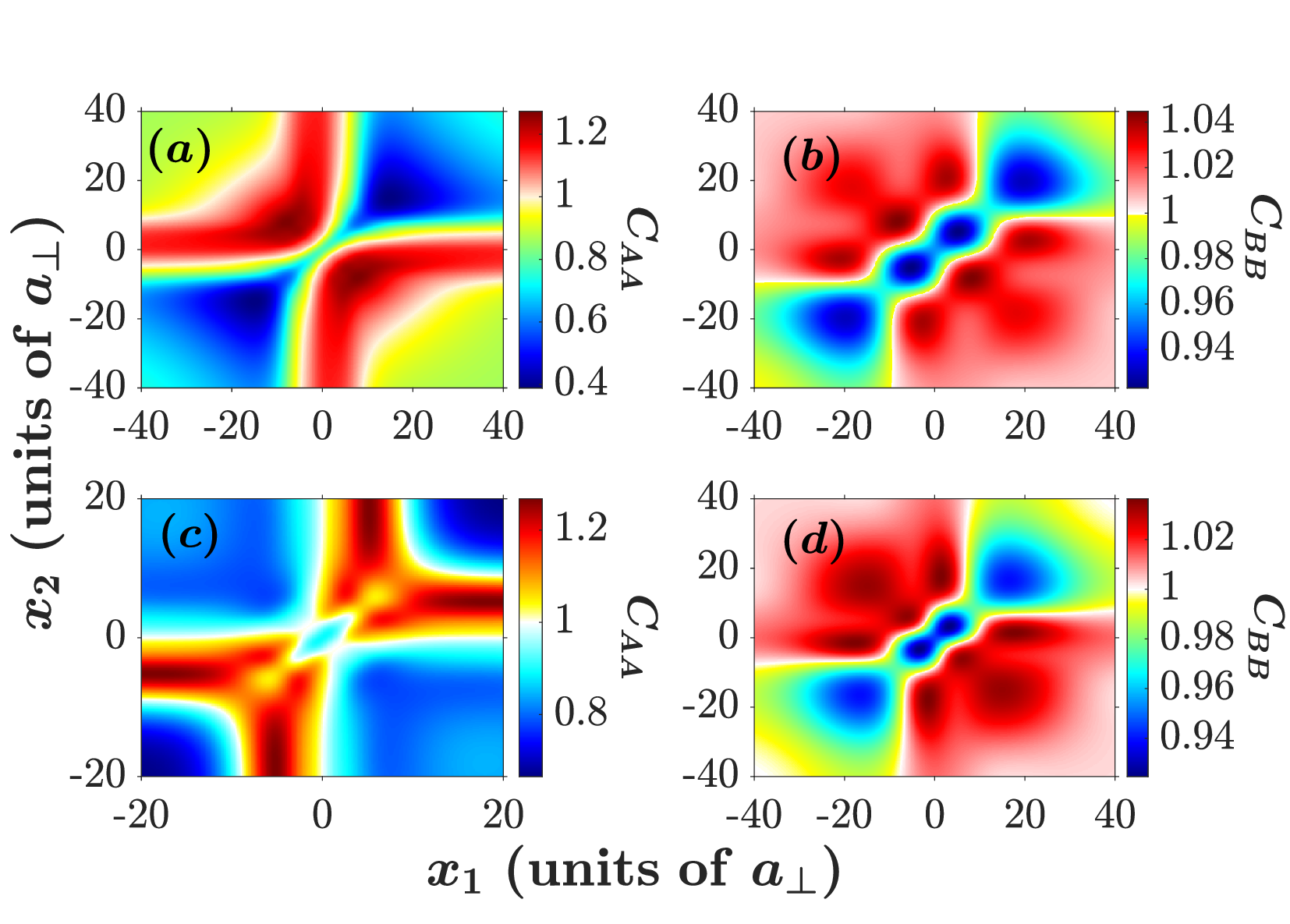}
\caption{Two-body coherence function of (a), (c) the minority component A and (b), (d) the majority component B for (a), (b) weak ($g_{\rm AB}=-0.02$) and (c), (d) strong ($ g_{\rm AB}= -0.142$) attraction. The remaining parameters are $g_A=g_B=0.1$, $N_A=2$, and $N_B=40$. The minority component exhibits a transition from droplet-like to soliton-like correlation patterns. Surprisingly, the majority component maintains its droplet character in all cases. }
\label{fig:GS2bd}
\end{figure} 

\section{Two-body droplet configurations}\label{sec:2bground}

\begin{figure*}[ht]
\centering
\includegraphics[width=0.98\textwidth]{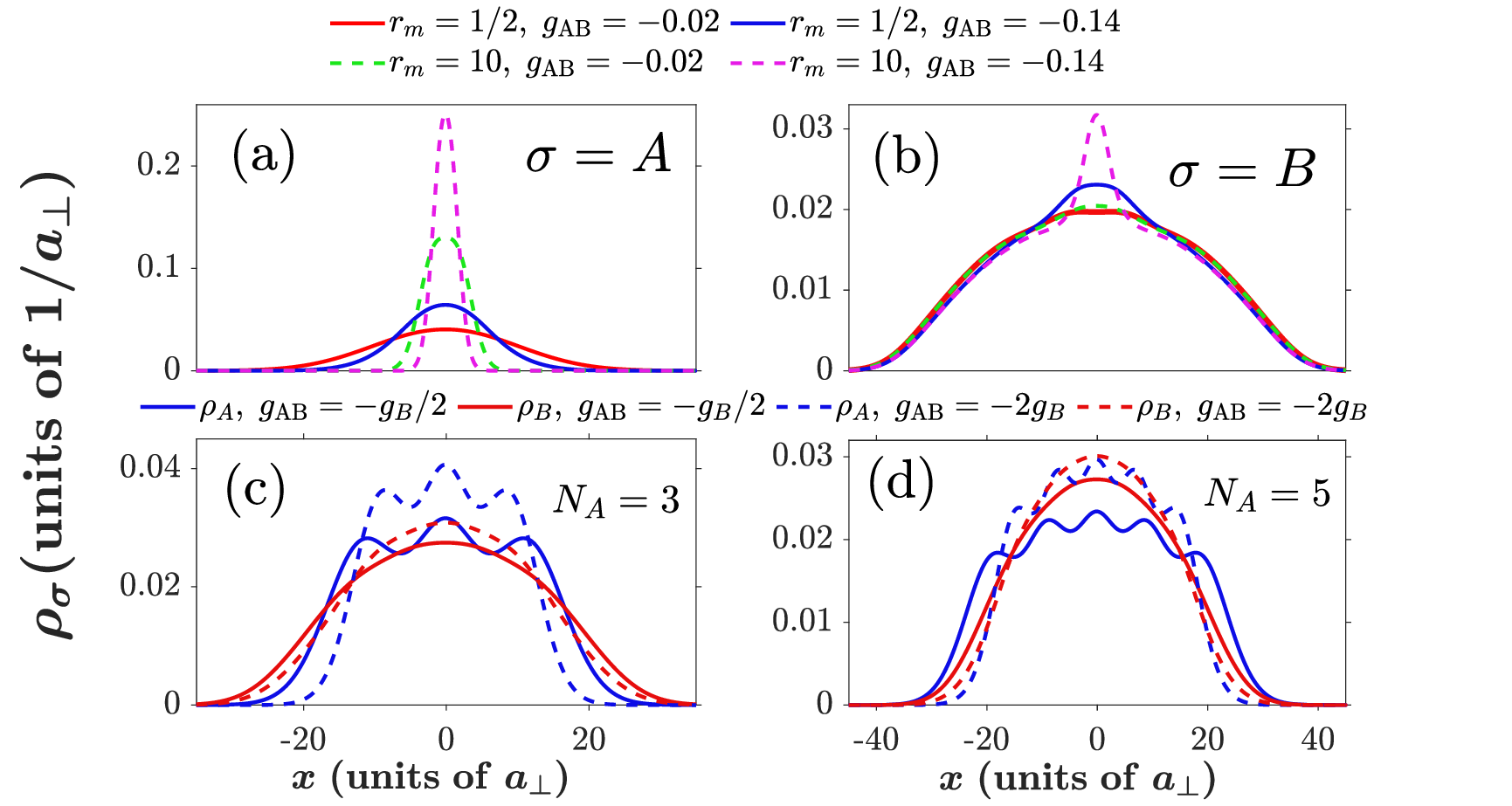}
\caption{Ground state droplet densities in heteronuclear Bose-Bose and Bose-Fermi mixtures within the many-body approach. 
Profiles of (a) $N_A=2$ bosons and (b) a  majority species with $N_B=40$ bosons for $g_A=g_B=0.1$ and varying mass ratio  $r_m = m_A/m_B$ and intercomponent attraction $g_{\rm AB}$ (see legend). 
Density configurations of (c) $N_A=3$ and (d) $N_A=5$ fermionic atoms within a majority species of $N_B=40$ bosons featuring $g_B=0.05$ for different intercomponent attractions $g_{\rm AB}$ (see legend).
Evidently, heavier minority bosons act against the FT droplet profile of the majority component, whilst droplet formation is suppressed in the presence of fermionic minority atoms.}
\label{fig:GSHetero}
\end{figure*} 

To further probe the superposition nature of the two-component droplet many-body states we examine the intracomponent two-body coherence functions $C_{\sigma\sigma}(x_1,x_2)=\rho^{(2)}_{\sigma\sigma}(x_1,x_2)/(\rho_\sigma(x_1)\rho_\sigma(x_2))$~\cite{mistakidis2018correlation,mistakidis2023few}. They are  defined in terms of the respective intracomponent two-body reduced densities
\begin{equation}\label{2bodyDens}
\begin{split}
    \rho^{(2)}_{\sigma\sigma}(x_1,x_2)= & \langle \Psi|\hat{\Psi}_\sigma ^\dagger (x_1) \hat{\Psi}_{\sigma} ^\dagger (x_2)  \hat{\Psi}_\sigma (x_1) \hat{\Psi}_{\sigma} (x_2) |\Psi \rangle,
\end{split}
\end{equation}
where $\hat{\Psi}_\sigma (x_i)$ [$\hat{\Psi}_\sigma^{\dagger} (x_i)$] refers to the bosonic field operator annihilating [creating] a $\sigma$-species atom at position $x_i$. $\rho^{(2)}_{\sigma\sigma'}(x_1,x_2)$ is the probability of simultaneously detecting a $\sigma$-species boson located at $x_1$ and another one at $x_2$~\cite{Sakmann2008Coher,Naraschewski1999Coher}. 
In this sense, two $\sigma$-species bosons show a bunching [anti-bunching] behavior if $C_{\sigma\sigma}(x_1,x_2;t)>1$ [$C_{\sigma \sigma}(x_1,x_2;t)<1$], and they are two-body uncorrelated for $C_{\sigma \sigma}(x_1,x_2;t)=1$.

The two-body correlation function, in the case of large particle imbalance i.e. $N_A/N_B = 1/20$, is provided in Fig.~\ref{fig:GS2bd} both for the majority and the minority components at two different intercomponent attractions. 
These are chosen such that the majority component has a FT shape and the minority exhibits a Gaussian profile. For weak attractions ($g_{\rm AB}=-0.02$) the components are almost  decoupled, while for stronger ones ($g_{\rm AB} =-0.142$) they are coupled and therefore the density of the majority is modulated within their overlap region. 
It can be seen that for increasing attraction the minority species features a transition from a two-body anti-correlated behavior (at the same position $x_1=x_2$) indicative of quantum droplets~\cite{ParisiGiorginiMonteCarlo,Mistakidis2021,Englezos2023} to a correlated pattern characteristic of soliton-like structures~\cite{MishmashPRL,KronkePRAMBsolitons,KronkePRAtwoBodySoliton,Katsimiga_2017} (compare in particular the main diagonal in Fig.~\ref{fig:GS2bd}(a) and (c)). 
This observation is further supported by the off-diagonal correlation behavior where for $g_{\rm AB}=-0.02$ ($g_{\rm AB} =-0.142$) two minority species atoms show an anti-bunching (bunching) tendency. 
This modified two-body correlation behavior is suggestive of a transition from a bright-droplet~\cite{AstrakharchikMalomed1DDynamics,katsimiga2023solitary} to a bright-soliton~\cite{pethick2008bose,Abdullaev2008} character for the minority component, as it is also indicated by Eq.~\eqref{bright} of our effective model. 
Elaborating further on the presence and properties of this transition is an intriguing prospect for future investigations. 
In contrast to the above, the majority component experiences an anti-bunching at the same location in both cases, see the  diagonal in Fig.~\ref{fig:GS2bd}(b) and (d), while two bosons placed symmetrically with respect to the FT are bunched. 
This correlation pattern further confirms our argument regarding the persistence of the underlying droplet character of these structures for large particle number ratios.

In contrast to the above behavior, for moderate particle number ratio, e.g. $N_A/N_B\approx 1/2$, both components undergo a progressive transition towards a localized solitonic structure characterized by a correlated  behavior upon increasing attraction (not shown). 
This is in agreement with the expectation that for systems close to particle balance, sufficiently strong intercomponent attraction gradually favors the collapse of the system~\cite{PetrovLowD,Stringari2016BEC,pethick2008bose}.

\section{imbalanced Droplets in heteronuclear  mixtures}\label{sec:hetero} 

Having described the peculiar ground-state droplet many-body configurations appearing in homonuclear  particle-imbalanced bosonic mixtures we next move to the investigation of heteronuclear (either bosonic or Bose-Fermi) ones. 
Specifically, we consider the same bosonic majority component as above but minority atoms composed of either a different bosonic element or fermionic isotope. 
Admittedly, the experimental preparation of such settings is more involved compared to the homonuclear mixtures. 
However, heteronuclear settings e.g. of $^{41}$K and $^{87}$Rb isotopes have been experimentally realized~\cite{FortHeteroExp} and importantly they offer the premise to unveil valuable insights on mechanisms that are absent in their single-component counterparts such as intercomponent  mixing, rich many-body phases and excitation processes~\cite{Mistakidis2021,Englezos2023}. 
Below, we solely rely on many-body ML-MCTDHX simulations since the eGPEs for heteronuclear 1D mixtures have not yet been constructed; rather they are known in 3D~\cite{AncilottoLocalDensity,FortHeteroExp}.

The case of $N_A=2$ heavy bosons immersed in a majority species of light $N_B=40$ bosons is presented in Fig.~\ref{fig:GSHetero}(a), (b) for various mass ratios $(r_m = m_A/m_B=1/2,\; 10)$ and intercomponent attractions $(g_{\rm AB}=-0.02,\; -0.14)$.  
As expected, due to their larger mass, the minority atoms experience gradually enhanced spatial localization for increasing mass ratio $m_A/m_B$ and fixed $g_{AB}$ or larger attraction $(g_{\rm AB})$ and constant $m_A/m_B$, see Fig.~\ref{fig:GSHetero}(a). 
As a consequence, the majority component shows progressively more pronounced spatial undulations in the vicinity of the minority species for either increasing mass ratio $m_A/m_B$ or attraction $g_{\rm AB}$. Accordingly, the width of the majority cloud slightly shrinks but it overall remains  approximately the same.   
In that light we can deduce that light minority species atoms coupled to the majority component  through weak attraction offer better candidates to  access the decoupled regime (see Eqs.~\eqref{eq: a}-\eqref{eq: b} in Sec.~\ref{sec:eGPEResults} and Fig.~\ref{fig:GSImb}(b)).
Recall, however, that in the presence of mass imbalance Eqs.~\eqref{eq: eGPEa}-\eqref{eq: eGPEb} are not valid. 
Interestingly, the phenomenology obtained for the mass-balanced and intercomponent particle imbalanced settings holds also for heteronuclear bosonic setups.

Next, we briefly address Bose-Fermi droplet settings where fermionic minority atoms, e.g. $N_A=3$ and $N_A=5$, are embedded within a majority species containing $N_B=30$ bosons, see Fig.~\ref{fig:GSHetero}(c), (d). 
Our analysis relies on the many-body ML-MCTDHX  approach~\cite{Mistakidis2019Fermions,Karpiuk2004,pethick2008bose}, in which the number states used for the expansion of the wave function given by Eq.~\eqref{permenantStates} become Slater-determinants of the $d_{\sigma}$ time-dependent variationally optimized single-particle functions, see also  Sec.~\ref{sec:Variational_Method}. 
As can be readily seen from Fig.~\ref{fig:GSHetero}(c) and (d), the shape of the Bose-Fermi mixture is strikingly different compared to the Bose-Bose one. 
Namely, the fermionic component tends to be equally or more delocalized than the bosonic majority species which is attributed to the Pauli exclusion principle. 
Also, as expected, for increasing intercomponent attraction both components become more localized. 
The energy per particle of the mixture is above the lowest trap  state. 
Clearly, the effective model presented in Sec.~\ref{sec:eGPEResults} was derived for a weakly interacting bosonic mixture and thus it is not  applicable for the Bose-Fermi setting. 

It has been argued that, at least for 3D systems~\cite{CuiSpinOrbitBoseFermi,GajdaBoseFermi,Wang_2020BoseFermi}, a highly imbalanced Bose-Fermi mixture with the bosonic component being the majority one, could accommodate  droplet structures. In~\cite{GajdaBoseFermi}, for example, it was explicated that a mixture of potassium $ ^{41}$K- $^{40}$K with densities $n_B\approx 10n_F$ (where $n_{B,[F]}$ is the bosonic [fermionic] density, normalized to the particle  number in the respective component) and interaction strength  ratio $|g_{BF}|/g_B \gtrapprox 0.25$ would result in a stable Bose-Fermi droplet in 3D free space.
To the best of our knowledge the possibility of Bose-Fermi droplet formation in 1D has not been investigated yet. 
Our results indicate that the relevant parameter region for the realization of Bose-Fermi droplets is significantly shifted in the 1D case as compared to the 3D one~\cite{CuiSpinOrbitBoseFermi,GajdaBoseFermi,Wang_2020BoseFermi}.
This could provide an interesting pathway for realizing 1D quantum Bose-Fermi droplets, since their 3D counterparts require large attractions and bosonic densities but also suffer from significant three-body recombination rates~\cite{CuiSpinOrbitBoseFermi,GajdaBoseFermi,Wang_2020BoseFermi}.
The latter being already suppressed in 1D, could potentially be further reduced if, as hinted by our results, 1D Bose-Fermi droplets prove to form in a parameter region with lesser three-body losses. 
Such a systematic study of 1D Bose-Fermi droplets is beyond the scope of our current work, however it would be intriguing to be pursued in the future.

We note in passing that in order to judge the degree of correlations in the Bose-Fermi mixture we have also inspected  the underlying orbital populations (not shown). 
It turns out that there is an increasing occupation of higher-lying species functions for larger attractions, while the  bosons mainly reside in the first orbital. 
This indicates an increase of the intercomponent entanglement, accompanied by minor intracomponent correlations for the bosonic species.
Hence, we find the opposite microscopic behavior for the Bose-Fermi mixture as compared to the Bose-Bose one, where the droplet-like states are characterized by significant intracomponent (anti-) correlations and relatively small intercomponent ones~\cite{ParisiMonteCarlo2019,ParisiGiorginiMonteCarlo,Mistakidis2021,Englezos2023}.
This is consistent with the absence of the signatures of the LHY phenomenology discussed above (see Sec.~\ref{sec:eGPEResults}) on the densities of the Bose-Fermi mixture in Fig.~\ref{fig:GSHetero}(c) and (d), since the LHY theory primarily accounts for the impact of intracomponent  correlations in the form of phonons~\cite{pethick2008bose,Stringari2016BEC,ParisiMonteCarlo2019,Hui2020LowD}.

\section{Summary and Perspectives}\label{sec:SummaryAndOutlook}

We have studied the formation of two-component bosonic droplet configurations with contact (intra-) inter-component (repulsion) attraction in the limit of large particle imbalance among the components. 
It is argued that the majority component can be arranged in a FT droplet shape exhibiting tunable in amplitude and spatial extent localized modulations in the vicinity of the minority atoms.  
These modulations become more pronounced for either increasing intercomponent attraction or intracomponent repulsion of the minority component as well as for larger mass of the latter. 
The intracomponent repulsion of the minority subsystem appears to have the stronger impact on the aforementioned undulations of the majority component. For instance, they vanish in the limit of non-interacting minority species. 
This qualitative behavior is analytically predicted via a reduction of the established eGPEs in the limit of large particle imbalance to a single-component effective model. 
It is further verified using many-body \textit{ab-initio} simulations within the ML-MCTDHX method. 

This many-body method enabled us to also address droplet formation in heteronuclear mixtures, where the corresponding 1D eGPEs are not available.
Specifically, for Bose-Bose settings it is found that heavier atoms in minority species enhance the localized undulations imprinted on the density of the majority species. 
Turning to Bose-Fermi systems we show that the FT signatures on the bosonic majority species vanish in the presence of fermions in the other component. 
This behavior supports the droplet suppression in Bose-Fermi mixtures.

Based on our results there is a multitude of future  research directions that can be pursued. 
A straightforward extension is to study the dynamical response of the identified droplet structures utilizing, for instance, quenches across the different phases in order to analyze the emergent pattern formation. 
The stability analysis of the two-component droplet configurations as it was done for the symmetric setting~\cite{katsimiga2023solitary} is highly desirable, while  considering spin-orbit coupling would introduce additional unstable modes~\cite{gangwar2023spectrum}. 
The characterization of such phases for strong interactions lying essentially beyond the validity of the eGPE would require to employ  sophisticated many-body methods, such as the ML-MCTDHX used herein or exact diagonalization~\cite{chergui2023superfluid}, for  capturing the underlying excitation spectrum and impact of thermal effects.

\section*{Acknowledgements} 

This work (P.S. and I.A.E.) has been funded by the Deutsche Forschungsgemeinschaft (DFG, German Research Foundation) - SFB 925 - project 170620586. S. I. M. acknowledges support from the NSF through a grant for ITAMP at Harvard University.

\appendix

\section{Comparison between the many-body and the eGPE predictions on the density profiles}\label{app:Comparison}

\begin{figure}[ht]
\centering
\includegraphics[width=0.47\textwidth]{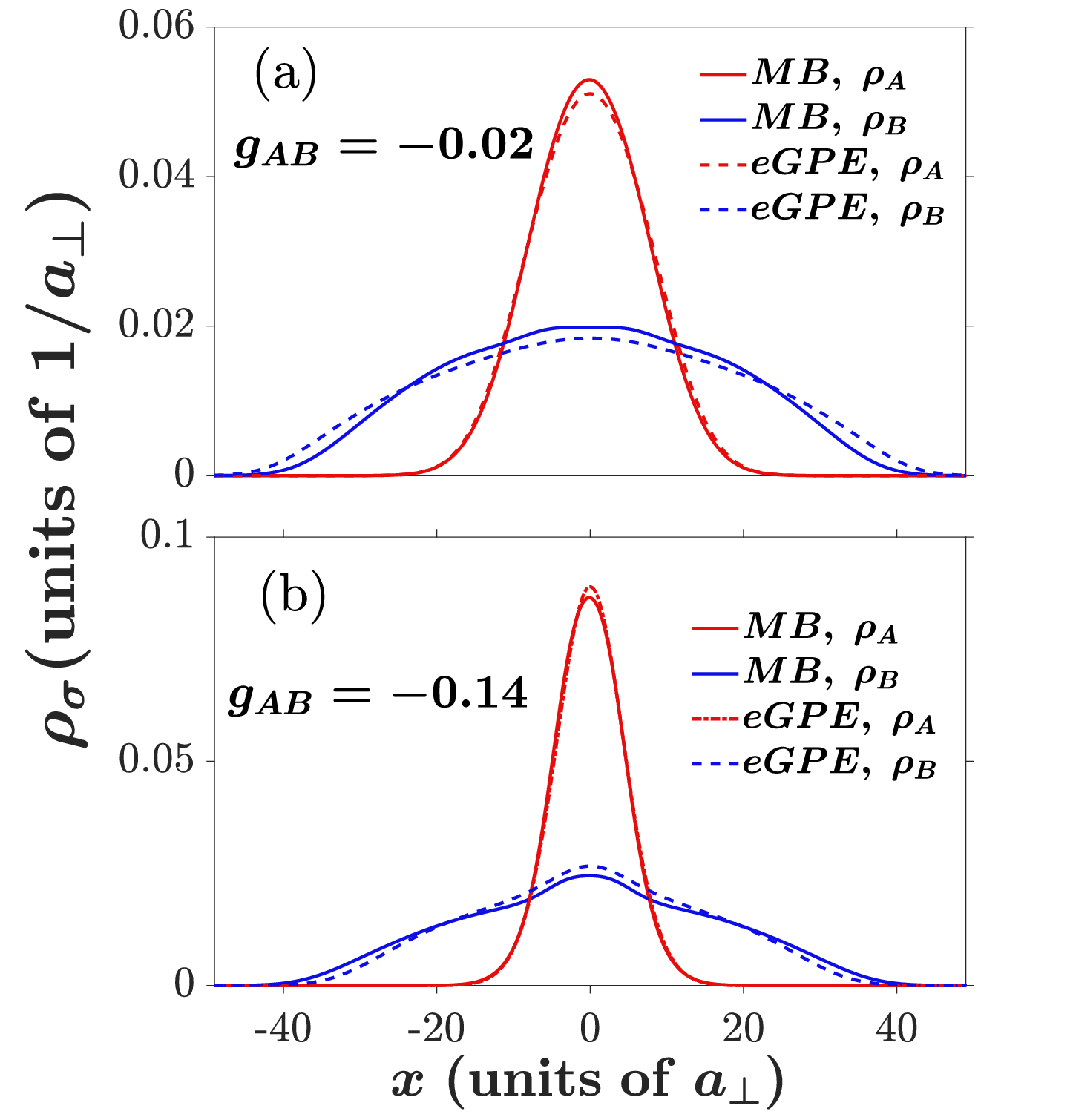}
\caption{Ground state droplet density profiles in homonuclear Bose-Bose harmonically confined ($\omega = 0.01$) mixtures within the many-body (MB) and the eGPE approaches (see legends). 
The mixture contains $N_A=4$, $N_B=40$ bosons and features  intercomponent attraction of strength (a) $g_{AB}=-0.02$, and (b) $g_{AB}=-0.14$.
In all cases the intracomponent repulsion is fixed to $g_A=g_B=0.1$.
The eGPE predictions are in qualitative agreement with the many-body results. 
Within the former approach a relatively smoother density profile occurs for the majority component but the FT behavior is absent.}
\label{fig:GSComp}
\end{figure} 

It is instructive to provide some additional insights on the ability of the LHY theory to capture the highly imbalanced two-component states discussed in the main text. 
For this reason, we present a short comparison on the single-particle density level obtained with the eGPEs and the {\it ab-initio} ML-MCTDHX method.

Paradigmatic ground state density profiles of the two component bosonic mixture with equal intraspecies repulsion ($g_A=g_B=0.1$) and $N_B=40$ ($N_A=4$) bosons in the majority (minority) component are illustrated in Fig.~\ref{fig:GSComp} for different interspecies attractions. 
The ground states are obtained numerically through the imaginary time propagation method applied either to the eGPEs [Eqs.~~\eqref{eGPE}(a), (b)] or the many-body ML-MCTDHX approach (see also Sec.~\ref{sec:Variational_Method}). 
The chosen parameter values are such that a transition from weak [Fig.~\ref{fig:GSComp}(a)] to stronger [Fig.~\ref{fig:GSComp}(b)] interspecies attraction takes place for $N_A=4$ similarly to the results shown in the main text, see also Fig.~\ref{fig:GSImb}(b). 
A careful comparison between the predictions of the many-body and the eGPEs reveals an adequate qualitative agreement of the resulting one-body spatial configurations of each component. 
Specifically, within the eGPE framework a slightly less (more) localized density profile for the majority component is obtained in the case of weak (stronger) attraction, as can be readily seen from Fig.~\ref{fig:GSComp}(a) [Fig.~\ref{fig:GSComp}(b)]. Furthermore, it is apparent that the density structures within the eGPEs are consistently smoother when compared to the corresponding many-body outcome, while the FT signatures present in the latter case are absent. 
This behavior is consistent with earlier predictions focusing on the symmetric mixture or equivalently single droplet case~\cite{ParisiGiorginiMonteCarlo,Englezos2023,mistakidis2023few} and attributing the emergent deviations to residual beyond-LHY correlations. 
Finally, it should be emphasized that despite the satisfactory agreement among the two approaches observed  on the single-particle density level, the many-body method allows to calculate higher-order observables such as correlation functions which are inaccessible with the eGPEs. 
At the level of these observables larger deviations are naturally expected especially in the course of the evolution.

\bibliographystyle{apsrev4-1}
\bibliography{ref_drops}	

\begin{thebibliography}{87}%
\makeatletter
\providecommand \@ifxundefined [1]{%
 \@ifx{#1\undefined}
}%
\providecommand \@ifnum [1]{%
 \ifnum #1\expandafter \@firstoftwo
 \else \expandafter \@secondoftwo
 \fi
}%
\providecommand \@ifx [1]{%
 \ifx #1\expandafter \@firstoftwo
 \else \expandafter \@secondoftwo
 \fi
}%
\providecommand \natexlab [1]{#1}%
\providecommand \enquote  [1]{``#1''}%
\providecommand \bibnamefont  [1]{#1}%
\providecommand \bibfnamefont [1]{#1}%
\providecommand \citenamefont [1]{#1}%
\providecommand \href@noop [0]{\@secondoftwo}%
\providecommand \href [0]{\begingroup \@sanitize@url \@href}%
\providecommand \@href[1]{\@@startlink{#1}\@@href}%
\providecommand \@@href[1]{\endgroup#1\@@endlink}%
\providecommand \@sanitize@url [0]{\catcode `\\12\catcode `\$12\catcode `\&12\catcode `\#12\catcode `\^12\catcode `\_12\catcode `\%12\relax}%
\providecommand \@@startlink[1]{}%
\providecommand \@@endlink[0]{}%
\providecommand \url  [0]{\begingroup\@sanitize@url \@url }%
\providecommand \@url [1]{\endgroup\@href {#1}{\urlprefix }}%
\providecommand \urlprefix  [0]{URL }%
\providecommand \Eprint [0]{\href }%
\providecommand \doibase [0]{http://dx.doi.org/}%
\providecommand \selectlanguage [0]{\@gobble}%
\providecommand \bibinfo  [0]{\@secondoftwo}%
\providecommand \bibfield  [0]{\@secondoftwo}%
\providecommand \translation [1]{[#1]}%
\providecommand \BibitemOpen [0]{}%
\providecommand \bibitemStop [0]{}%
\providecommand \bibitemNoStop [0]{.\EOS\space}%
\providecommand \EOS [0]{\spacefactor3000\relax}%
\providecommand \BibitemShut  [1]{\csname bibitem#1\endcsname}%
\let\auto@bib@innerbib\@empty
\bibitem [{\citenamefont {Bloch}\ \emph {et~al.}(2012)\citenamefont {Bloch}, \citenamefont {Dalibard},\ and\ \citenamefont {Nascimbène}}]{BlochNature2012}%
  \BibitemOpen
  \bibfield  {author} {\bibinfo {author} {\bibfnamefont {I.}~\bibnamefont {Bloch}}, \bibinfo {author} {\bibfnamefont {J.}~\bibnamefont {Dalibard}}, \ and\ \bibinfo {author} {\bibfnamefont {S.}~\bibnamefont {Nascimbène}},\ }\href {\doibase 10.1038/nphys2259} {\bibfield  {journal} {\bibinfo  {journal} {Nature Phys.}\ }\textbf {\bibinfo {volume} {8}},\ \bibinfo {pages} {267} (\bibinfo {year} {2012})}\BibitemShut {NoStop}%
\bibitem [{\citenamefont {Petrov}(2015)}]{Petrov2015}%
  \BibitemOpen
  \bibfield  {author} {\bibinfo {author} {\bibfnamefont {D.~S.}\ \bibnamefont {Petrov}},\ }\href {\doibase 10.1103/PhysRevLett.115.155302} {\bibfield  {journal} {\bibinfo  {journal} {Phys. Rev. Lett.}\ }\textbf {\bibinfo {volume} {115}},\ \bibinfo {pages} {155302} (\bibinfo {year} {2015})}\BibitemShut {NoStop}%
\bibitem [{\citenamefont {Ferrier-Barbut}\ \emph {et~al.}(2016)\citenamefont {Ferrier-Barbut}, \citenamefont {Kadau}, \citenamefont {Schmitt}, \citenamefont {Wenzel},\ and\ \citenamefont {Pfau}}]{KadauDropExp}%
  \BibitemOpen
  \bibfield  {author} {\bibinfo {author} {\bibfnamefont {I.}~\bibnamefont {Ferrier-Barbut}}, \bibinfo {author} {\bibfnamefont {H.}~\bibnamefont {Kadau}}, \bibinfo {author} {\bibfnamefont {M.}~\bibnamefont {Schmitt}}, \bibinfo {author} {\bibfnamefont {M.}~\bibnamefont {Wenzel}}, \ and\ \bibinfo {author} {\bibfnamefont {T.}~\bibnamefont {Pfau}},\ }\href {\doibase 10.1103/PhysRevLett.116.215301} {\bibfield  {journal} {\bibinfo  {journal} {Phys. Rev. Lett.}\ }\textbf {\bibinfo {volume} {116}},\ \bibinfo {pages} {215301} (\bibinfo {year} {2016})}\BibitemShut {NoStop}%
\bibitem [{\citenamefont {Böttcher}\ \emph {et~al.}(2020)\citenamefont {Böttcher}, \citenamefont {Schmidt}, \citenamefont {Hertkorn}, \citenamefont {Ng}, \citenamefont {Graham}, \citenamefont {Guo}, \citenamefont {Langen},\ and\ \citenamefont {Pfau}}]{PfauReview}%
  \BibitemOpen
  \bibfield  {author} {\bibinfo {author} {\bibfnamefont {F.}~\bibnamefont {Böttcher}}, \bibinfo {author} {\bibfnamefont {J.-N.}\ \bibnamefont {Schmidt}}, \bibinfo {author} {\bibfnamefont {J.}~\bibnamefont {Hertkorn}}, \bibinfo {author} {\bibfnamefont {K.~S.~H.}\ \bibnamefont {Ng}}, \bibinfo {author} {\bibfnamefont {S.~D.}\ \bibnamefont {Graham}}, \bibinfo {author} {\bibfnamefont {M.}~\bibnamefont {Guo}}, \bibinfo {author} {\bibfnamefont {T.}~\bibnamefont {Langen}}, \ and\ \bibinfo {author} {\bibfnamefont {T.}~\bibnamefont {Pfau}},\ }\href {\doibase 10.1088/1361-6633/abc9ab} {\bibfield  {journal} {\bibinfo  {journal} {Rep. Progr. Phys.}\ }\textbf {\bibinfo {volume} {84}},\ \bibinfo {pages} {012403} (\bibinfo {year} {2020})}\BibitemShut {NoStop}%
\bibitem [{\citenamefont {Luo}\ \emph {et~al.}(2020)\citenamefont {Luo}, \citenamefont {Pang}, \citenamefont {Liu}, \citenamefont {Li},\ and\ \citenamefont {Malomed}}]{MalomedLuoReview}%
  \BibitemOpen
  \bibfield  {author} {\bibinfo {author} {\bibfnamefont {Z.-H.}\ \bibnamefont {Luo}}, \bibinfo {author} {\bibfnamefont {W.}~\bibnamefont {Pang}}, \bibinfo {author} {\bibfnamefont {B.}~\bibnamefont {Liu}}, \bibinfo {author} {\bibfnamefont {Y.-Y.}\ \bibnamefont {Li}}, \ and\ \bibinfo {author} {\bibfnamefont {B.~A.}\ \bibnamefont {Malomed}},\ }\href {\doibase 10.1007/s11467-020-1020-2} {\bibfield  {journal} {\bibinfo  {journal} {Front. Phys.}\ }\textbf {\bibinfo {volume} {16}},\ \bibinfo {pages} {32201} (\bibinfo {year} {2020})}\BibitemShut {NoStop}%
\bibitem [{\citenamefont {Malomed}(2021)}]{MalomedReview}%
  \BibitemOpen
  \bibfield  {author} {\bibinfo {author} {\bibfnamefont {B.~A.}\ \bibnamefont {Malomed}},\ }\href {\doibase 10.1007/s11467-020-1024-y} {\bibfield  {journal} {\bibinfo  {journal} {Front. Phys.}\ }\textbf {\bibinfo {volume} {16}},\ \bibinfo {eid} {22504} (\bibinfo {year} {2021})}\BibitemShut {NoStop}%
\bibitem [{\citenamefont {Massignan}\ \emph {et~al.}(2014)\citenamefont {Massignan}, \citenamefont {Zaccanti},\ and\ \citenamefont {Bruun}}]{PolaronsBruun}%
  \BibitemOpen
  \bibfield  {author} {\bibinfo {author} {\bibfnamefont {P.}~\bibnamefont {Massignan}}, \bibinfo {author} {\bibfnamefont {M.}~\bibnamefont {Zaccanti}}, \ and\ \bibinfo {author} {\bibfnamefont {G.~M.}\ \bibnamefont {Bruun}},\ }\href {\doibase 10.1088/0034-4885/77/3/034401} {\bibfield  {journal} {\bibinfo  {journal} {Rep. Prog. Phys.}\ }\textbf {\bibinfo {volume} {77}},\ \bibinfo {pages} {034401} (\bibinfo {year} {2014})}\BibitemShut {NoStop}%
\bibitem [{\citenamefont {Schmidt}\ \emph {et~al.}(2018)\citenamefont {Schmidt}, \citenamefont {Knap}, \citenamefont {Ivanov}, \citenamefont {You}, \citenamefont {Cetina},\ and\ \citenamefont {Demler}}]{PolaronsDemler}%
  \BibitemOpen
  \bibfield  {author} {\bibinfo {author} {\bibfnamefont {R.}~\bibnamefont {Schmidt}}, \bibinfo {author} {\bibfnamefont {M.}~\bibnamefont {Knap}}, \bibinfo {author} {\bibfnamefont {D.~A.}\ \bibnamefont {Ivanov}}, \bibinfo {author} {\bibfnamefont {J.-S.}\ \bibnamefont {You}}, \bibinfo {author} {\bibfnamefont {M.}~\bibnamefont {Cetina}}, \ and\ \bibinfo {author} {\bibfnamefont {E.}~\bibnamefont {Demler}},\ }\href {\doibase 10.1088/1361-6633/aa9593} {\bibfield  {journal} {\bibinfo  {journal} {Rep. Prog. Phys.}\ }\textbf {\bibinfo {volume} {81}},\ \bibinfo {pages} {024401} (\bibinfo {year} {2018})}\BibitemShut {NoStop}%
\bibitem [{\citenamefont {Fukuhara}\ \emph {et~al.}(2013)\citenamefont {Fukuhara}, \citenamefont {Kantian}, \citenamefont {Endres}, \citenamefont {Cheneau}, \citenamefont {Schauß}, \citenamefont {Hild}, \citenamefont {Bellem}, \citenamefont {Schollwöck}, \citenamefont {Giamarchi}, \citenamefont {Gross}, \citenamefont {Bloch},\ and\ \citenamefont {Kuhr}}]{BosePolaronExp1}%
  \BibitemOpen
  \bibfield  {author} {\bibinfo {author} {\bibfnamefont {T.}~\bibnamefont {Fukuhara}}, \bibinfo {author} {\bibfnamefont {A.}~\bibnamefont {Kantian}}, \bibinfo {author} {\bibfnamefont {M.}~\bibnamefont {Endres}}, \bibinfo {author} {\bibfnamefont {M.}~\bibnamefont {Cheneau}}, \bibinfo {author} {\bibfnamefont {P.}~\bibnamefont {Schauß}}, \bibinfo {author} {\bibfnamefont {S.}~\bibnamefont {Hild}}, \bibinfo {author} {\bibfnamefont {D.}~\bibnamefont {Bellem}}, \bibinfo {author} {\bibfnamefont {U.}~\bibnamefont {Schollwöck}}, \bibinfo {author} {\bibfnamefont {T.}~\bibnamefont {Giamarchi}}, \bibinfo {author} {\bibfnamefont {C.}~\bibnamefont {Gross}}, \bibinfo {author} {\bibfnamefont {I.}~\bibnamefont {Bloch}}, \ and\ \bibinfo {author} {\bibfnamefont {S.}~\bibnamefont {Kuhr}},\ }\href {\doibase 10.1038/nphys2561} {\bibfield  {journal} {\bibinfo  {journal} {Nature Physics}\ }\textbf {\bibinfo {volume} {9}},\ \bibinfo {pages} {235} (\bibinfo {year} {2013})}\BibitemShut {NoStop}%
\bibitem [{\citenamefont {Catani}\ \emph {et~al.}(2009)\citenamefont {Catani}, \citenamefont {Barontini}, \citenamefont {Lamporesi}, \citenamefont {Rabatti}, \citenamefont {Thalhammer}, \citenamefont {Minardi}, \citenamefont {Stringari},\ and\ \citenamefont {Inguscio}}]{BosePolaronExp2}%
  \BibitemOpen
  \bibfield  {author} {\bibinfo {author} {\bibfnamefont {J.}~\bibnamefont {Catani}}, \bibinfo {author} {\bibfnamefont {G.}~\bibnamefont {Barontini}}, \bibinfo {author} {\bibfnamefont {G.}~\bibnamefont {Lamporesi}}, \bibinfo {author} {\bibfnamefont {F.}~\bibnamefont {Rabatti}}, \bibinfo {author} {\bibfnamefont {G.}~\bibnamefont {Thalhammer}}, \bibinfo {author} {\bibfnamefont {F.}~\bibnamefont {Minardi}}, \bibinfo {author} {\bibfnamefont {S.}~\bibnamefont {Stringari}}, \ and\ \bibinfo {author} {\bibfnamefont {M.}~\bibnamefont {Inguscio}},\ }\href {\doibase 10.1103/PhysRevLett.103.140401} {\bibfield  {journal} {\bibinfo  {journal} {Phys. Rev. Lett.}\ }\textbf {\bibinfo {volume} {103}},\ \bibinfo {pages} {140401} (\bibinfo {year} {2009})}\BibitemShut {NoStop}%
\bibitem [{\citenamefont {J\o{}rgensen}\ \emph {et~al.}(2016)\citenamefont {J\o{}rgensen}, \citenamefont {Wacker}, \citenamefont {Skalmstang}, \citenamefont {Parish}, \citenamefont {Levinsen}, \citenamefont {Christensen}, \citenamefont {Bruun},\ and\ \citenamefont {Arlt}}]{BosePolaronExp3}%
  \BibitemOpen
  \bibfield  {author} {\bibinfo {author} {\bibfnamefont {N.~B.}\ \bibnamefont {J\o{}rgensen}}, \bibinfo {author} {\bibfnamefont {L.}~\bibnamefont {Wacker}}, \bibinfo {author} {\bibfnamefont {K.~T.}\ \bibnamefont {Skalmstang}}, \bibinfo {author} {\bibfnamefont {M.~M.}\ \bibnamefont {Parish}}, \bibinfo {author} {\bibfnamefont {J.}~\bibnamefont {Levinsen}}, \bibinfo {author} {\bibfnamefont {R.~S.}\ \bibnamefont {Christensen}}, \bibinfo {author} {\bibfnamefont {G.~M.}\ \bibnamefont {Bruun}}, \ and\ \bibinfo {author} {\bibfnamefont {J.~J.}\ \bibnamefont {Arlt}},\ }\href {\doibase 10.1103/PhysRevLett.117.055302} {\bibfield  {journal} {\bibinfo  {journal} {Phys. Rev. Lett.}\ }\textbf {\bibinfo {volume} {117}},\ \bibinfo {pages} {055302} (\bibinfo {year} {2016})}\BibitemShut {NoStop}%
\bibitem [{\citenamefont {Hu}\ \emph {et~al.}(2016)\citenamefont {Hu}, \citenamefont {Van~de Graaff}, \citenamefont {Kedar}, \citenamefont {Corson}, \citenamefont {Cornell},\ and\ \citenamefont {Jin}}]{BosePolaronExp4}%
  \BibitemOpen
  \bibfield  {author} {\bibinfo {author} {\bibfnamefont {M.-G.}\ \bibnamefont {Hu}}, \bibinfo {author} {\bibfnamefont {M.~J.}\ \bibnamefont {Van~de Graaff}}, \bibinfo {author} {\bibfnamefont {D.}~\bibnamefont {Kedar}}, \bibinfo {author} {\bibfnamefont {J.~P.}\ \bibnamefont {Corson}}, \bibinfo {author} {\bibfnamefont {E.~A.}\ \bibnamefont {Cornell}}, \ and\ \bibinfo {author} {\bibfnamefont {D.~S.}\ \bibnamefont {Jin}},\ }\href {\doibase 10.1103/PhysRevLett.117.055301} {\bibfield  {journal} {\bibinfo  {journal} {Phys. Rev. Lett.}\ }\textbf {\bibinfo {volume} {117}},\ \bibinfo {pages} {055301} (\bibinfo {year} {2016})}\BibitemShut {NoStop}%
\bibitem [{\citenamefont {Yan}\ \emph {et~al.}(2020)\citenamefont {Yan}, \citenamefont {Ni}, \citenamefont {Robens},\ and\ \citenamefont {Zwierlein}}]{BosePolaronExp5}%
  \BibitemOpen
  \bibfield  {author} {\bibinfo {author} {\bibfnamefont {Z.~Z.}\ \bibnamefont {Yan}}, \bibinfo {author} {\bibfnamefont {Y.}~\bibnamefont {Ni}}, \bibinfo {author} {\bibfnamefont {C.}~\bibnamefont {Robens}}, \ and\ \bibinfo {author} {\bibfnamefont {M.~W.}\ \bibnamefont {Zwierlein}},\ }\href {\doibase 10.1126/science.aax5850} {\bibfield  {journal} {\bibinfo  {journal} {Science}\ }\textbf {\bibinfo {volume} {368}},\ \bibinfo {pages} {190} (\bibinfo {year} {2020})}\BibitemShut {NoStop}%
\bibitem [{\citenamefont {Kohstall}\ \emph {et~al.}(2012)\citenamefont {Kohstall}, \citenamefont {Zaccanti}, \citenamefont {Jag}, \citenamefont {Trenkwalder}, \citenamefont {Massignan}, \citenamefont {Bruun}, \citenamefont {Schreck},\ and\ \citenamefont {Grimm}}]{FermiPolaronExp1}%
  \BibitemOpen
  \bibfield  {author} {\bibinfo {author} {\bibfnamefont {C.}~\bibnamefont {Kohstall}}, \bibinfo {author} {\bibfnamefont {M.}~\bibnamefont {Zaccanti}}, \bibinfo {author} {\bibfnamefont {M.}~\bibnamefont {Jag}}, \bibinfo {author} {\bibfnamefont {A.}~\bibnamefont {Trenkwalder}}, \bibinfo {author} {\bibfnamefont {P.}~\bibnamefont {Massignan}}, \bibinfo {author} {\bibfnamefont {G.~M.}\ \bibnamefont {Bruun}}, \bibinfo {author} {\bibfnamefont {F.}~\bibnamefont {Schreck}}, \ and\ \bibinfo {author} {\bibfnamefont {R.}~\bibnamefont {Grimm}},\ }\href {\doibase 10.1038/nature11065} {\bibfield  {journal} {\bibinfo  {journal} {Nature}\ }\textbf {\bibinfo {volume} {485}},\ \bibinfo {pages} {615} (\bibinfo {year} {2012})}\BibitemShut {NoStop}%
\bibitem [{\citenamefont {Scazza}\ \emph {et~al.}(2017)\citenamefont {Scazza}, \citenamefont {Valtolina}, \citenamefont {Massignan}, \citenamefont {Recati}, \citenamefont {Amico}, \citenamefont {Burchianti}, \citenamefont {Fort}, \citenamefont {Inguscio}, \citenamefont {Zaccanti},\ and\ \citenamefont {Roati}}]{FermiPolaronExp2}%
  \BibitemOpen
  \bibfield  {author} {\bibinfo {author} {\bibfnamefont {F.}~\bibnamefont {Scazza}}, \bibinfo {author} {\bibfnamefont {G.}~\bibnamefont {Valtolina}}, \bibinfo {author} {\bibfnamefont {P.}~\bibnamefont {Massignan}}, \bibinfo {author} {\bibfnamefont {A.}~\bibnamefont {Recati}}, \bibinfo {author} {\bibfnamefont {A.}~\bibnamefont {Amico}}, \bibinfo {author} {\bibfnamefont {A.}~\bibnamefont {Burchianti}}, \bibinfo {author} {\bibfnamefont {C.}~\bibnamefont {Fort}}, \bibinfo {author} {\bibfnamefont {M.}~\bibnamefont {Inguscio}}, \bibinfo {author} {\bibfnamefont {M.}~\bibnamefont {Zaccanti}}, \ and\ \bibinfo {author} {\bibfnamefont {G.}~\bibnamefont {Roati}},\ }\href {\doibase 10.1103/PhysRevLett.118.083602} {\bibfield  {journal} {\bibinfo  {journal} {Phys. Rev. Lett.}\ }\textbf {\bibinfo {volume} {118}},\ \bibinfo {pages} {083602} (\bibinfo {year} {2017})}\BibitemShut {NoStop}%
\bibitem [{\citenamefont {Koschorreck}\ \emph {et~al.}(2012)\citenamefont {Koschorreck}, \citenamefont {Pertot}, \citenamefont {Vogt}, \citenamefont {Fröhlich}, \citenamefont {Feld},\ and\ \citenamefont {Köhl}}]{FermiPolaronExp3}%
  \BibitemOpen
  \bibfield  {author} {\bibinfo {author} {\bibfnamefont {M.}~\bibnamefont {Koschorreck}}, \bibinfo {author} {\bibfnamefont {D.}~\bibnamefont {Pertot}}, \bibinfo {author} {\bibfnamefont {E.}~\bibnamefont {Vogt}}, \bibinfo {author} {\bibfnamefont {B.}~\bibnamefont {Fröhlich}}, \bibinfo {author} {\bibfnamefont {M.}~\bibnamefont {Feld}}, \ and\ \bibinfo {author} {\bibfnamefont {M.}~\bibnamefont {Köhl}},\ }\href {\doibase 10.1038/nature11151} {\bibfield  {journal} {\bibinfo  {journal} {Nature}\ }\textbf {\bibinfo {volume} {485}},\ \bibinfo {pages} {619} (\bibinfo {year} {2012})}\BibitemShut {NoStop}%
\bibitem [{\citenamefont {Demler}\ and\ \citenamefont {Grusdt}(2015)}]{BosePolaronDemler}%
  \BibitemOpen
  \bibfield  {author} {\bibinfo {author} {\bibfnamefont {E.}~\bibnamefont {Demler}}\ and\ \bibinfo {author} {\bibfnamefont {F.}~\bibnamefont {Grusdt}},\ }\href {https://doi.org/10.48550/arXiv.1510.04934} {\bibfield  {journal} {\bibinfo  {journal} {arXiv:1510.04934}\ } (\bibinfo {year} {2015})}\BibitemShut {NoStop}%
\bibitem [{\citenamefont {Rath}\ and\ \citenamefont {Schmidt}(2013)}]{BosePolaronFieldTheorySchidt}%
  \BibitemOpen
  \bibfield  {author} {\bibinfo {author} {\bibfnamefont {S.~P.}\ \bibnamefont {Rath}}\ and\ \bibinfo {author} {\bibfnamefont {R.}~\bibnamefont {Schmidt}},\ }\href {\doibase 10.1103/PhysRevA.88.053632} {\bibfield  {journal} {\bibinfo  {journal} {Phys. Rev. A}\ }\textbf {\bibinfo {volume} {88}},\ \bibinfo {pages} {053632} (\bibinfo {year} {2013})}\BibitemShut {NoStop}%
\bibitem [{\citenamefont {Bruderer}\ \emph {et~al.}(2007)\citenamefont {Bruderer}, \citenamefont {Klein}, \citenamefont {Clark},\ and\ \citenamefont {Jaksch}}]{PolaronsJaksch}%
  \BibitemOpen
  \bibfield  {author} {\bibinfo {author} {\bibfnamefont {M.}~\bibnamefont {Bruderer}}, \bibinfo {author} {\bibfnamefont {A.}~\bibnamefont {Klein}}, \bibinfo {author} {\bibfnamefont {S.~R.}\ \bibnamefont {Clark}}, \ and\ \bibinfo {author} {\bibfnamefont {D.}~\bibnamefont {Jaksch}},\ }\href {\doibase 10.1103/PhysRevA.76.011605} {\bibfield  {journal} {\bibinfo  {journal} {Phys. Rev. A}\ }\textbf {\bibinfo {volume} {76}},\ \bibinfo {pages} {011605} (\bibinfo {year} {2007})}\BibitemShut {NoStop}%
\bibitem [{\citenamefont {Dehkharghani}\ \emph {et~al.}(2018)\citenamefont {Dehkharghani}, \citenamefont {Volosniev},\ and\ \citenamefont {Zinner}}]{PolaronsZinner}%
  \BibitemOpen
  \bibfield  {author} {\bibinfo {author} {\bibfnamefont {A.~S.}\ \bibnamefont {Dehkharghani}}, \bibinfo {author} {\bibfnamefont {A.~G.}\ \bibnamefont {Volosniev}}, \ and\ \bibinfo {author} {\bibfnamefont {N.~T.}\ \bibnamefont {Zinner}},\ }\href {\doibase 10.1103/PhysRevLett.121.080405} {\bibfield  {journal} {\bibinfo  {journal} {Phys. Rev. Lett.}\ }\textbf {\bibinfo {volume} {121}},\ \bibinfo {pages} {080405} (\bibinfo {year} {2018})}\BibitemShut {NoStop}%
\bibitem [{\citenamefont {Ardila}\ and\ \citenamefont {Giorgini}(2015)}]{PolaronsGiorgini}%
  \BibitemOpen
  \bibfield  {author} {\bibinfo {author} {\bibfnamefont {L.~A. P.~n.}\ \bibnamefont {Ardila}}\ and\ \bibinfo {author} {\bibfnamefont {S.}~\bibnamefont {Giorgini}},\ }\href {\doibase 10.1103/PhysRevA.92.033612} {\bibfield  {journal} {\bibinfo  {journal} {Phys. Rev. A}\ }\textbf {\bibinfo {volume} {92}},\ \bibinfo {pages} {033612} (\bibinfo {year} {2015})}\BibitemShut {NoStop}%
\bibitem [{\citenamefont {Grusdt}\ \emph {et~al.}(2017)\citenamefont {Grusdt}, \citenamefont {Schmidt}, \citenamefont {Shchadilova},\ and\ \citenamefont {Demler}}]{PolaronsDemler2017}%
  \BibitemOpen
  \bibfield  {author} {\bibinfo {author} {\bibfnamefont {F.}~\bibnamefont {Grusdt}}, \bibinfo {author} {\bibfnamefont {R.}~\bibnamefont {Schmidt}}, \bibinfo {author} {\bibfnamefont {Y.~E.}\ \bibnamefont {Shchadilova}}, \ and\ \bibinfo {author} {\bibfnamefont {E.}~\bibnamefont {Demler}},\ }\href {\doibase 10.1103/PhysRevA.96.013607} {\bibfield  {journal} {\bibinfo  {journal} {Phys. Rev. A}\ }\textbf {\bibinfo {volume} {96}},\ \bibinfo {pages} {013607} (\bibinfo {year} {2017})}\BibitemShut {NoStop}%
\bibitem [{\citenamefont {Ichmoukhamedov}\ and\ \citenamefont {Tempere}(2019)}]{PolaronsTempere}%
  \BibitemOpen
  \bibfield  {author} {\bibinfo {author} {\bibfnamefont {T.}~\bibnamefont {Ichmoukhamedov}}\ and\ \bibinfo {author} {\bibfnamefont {J.}~\bibnamefont {Tempere}},\ }\href {\doibase 10.1103/PhysRevA.100.043605} {\bibfield  {journal} {\bibinfo  {journal} {Phys. Rev. A}\ }\textbf {\bibinfo {volume} {100}},\ \bibinfo {pages} {043605} (\bibinfo {year} {2019})}\BibitemShut {NoStop}%
\bibitem [{\citenamefont {Volosniev}\ \emph {et~al.}(2015)\citenamefont {Volosniev}, \citenamefont {Hammer},\ and\ \citenamefont {Zinner}}]{PolaronDynamics1}%
  \BibitemOpen
  \bibfield  {author} {\bibinfo {author} {\bibfnamefont {A.~G.}\ \bibnamefont {Volosniev}}, \bibinfo {author} {\bibfnamefont {H.-W.}\ \bibnamefont {Hammer}}, \ and\ \bibinfo {author} {\bibfnamefont {N.~T.}\ \bibnamefont {Zinner}},\ }\href {\doibase 10.1103/PhysRevA.92.023623} {\bibfield  {journal} {\bibinfo  {journal} {Phys. Rev. A}\ }\textbf {\bibinfo {volume} {92}},\ \bibinfo {pages} {023623} (\bibinfo {year} {2015})}\BibitemShut {NoStop}%
\bibitem [{\citenamefont {Mistakidis}\ \emph {et~al.}(2019{\natexlab{a}})\citenamefont {Mistakidis}, \citenamefont {Katsimiga}, \citenamefont {Koutentakis}, \citenamefont {Busch},\ and\ \citenamefont {Schmelcher}}]{PolaronDynamics2}%
  \BibitemOpen
  \bibfield  {author} {\bibinfo {author} {\bibfnamefont {S.~I.}\ \bibnamefont {Mistakidis}}, \bibinfo {author} {\bibfnamefont {G.~C.}\ \bibnamefont {Katsimiga}}, \bibinfo {author} {\bibfnamefont {G.~M.}\ \bibnamefont {Koutentakis}}, \bibinfo {author} {\bibfnamefont {T.}~\bibnamefont {Busch}}, \ and\ \bibinfo {author} {\bibfnamefont {P.}~\bibnamefont {Schmelcher}},\ }\href {\doibase 10.1103/PhysRevLett.122.183001} {\bibfield  {journal} {\bibinfo  {journal} {Phys. Rev. Lett.}\ }\textbf {\bibinfo {volume} {122}},\ \bibinfo {pages} {183001} (\bibinfo {year} {2019}{\natexlab{a}})}\BibitemShut {NoStop}%
\bibitem [{\citenamefont {Mistakidis}\ \emph {et~al.}(2019{\natexlab{b}})\citenamefont {Mistakidis}, \citenamefont {Volosniev}, \citenamefont {Zinner},\ and\ \citenamefont {Schmelcher}}]{PolaronDynamics3}%
  \BibitemOpen
  \bibfield  {author} {\bibinfo {author} {\bibfnamefont {S.~I.}\ \bibnamefont {Mistakidis}}, \bibinfo {author} {\bibfnamefont {A.~G.}\ \bibnamefont {Volosniev}}, \bibinfo {author} {\bibfnamefont {N.~T.}\ \bibnamefont {Zinner}}, \ and\ \bibinfo {author} {\bibfnamefont {P.}~\bibnamefont {Schmelcher}},\ }\href {\doibase 10.1103/PhysRevA.100.013619} {\bibfield  {journal} {\bibinfo  {journal} {Phys. Rev. A}\ }\textbf {\bibinfo {volume} {100}},\ \bibinfo {pages} {013619} (\bibinfo {year} {2019}{\natexlab{b}})}\BibitemShut {NoStop}%
\bibitem [{\citenamefont {Grusdt}\ \emph {et~al.}(2018)\citenamefont {Grusdt}, \citenamefont {Seetharam}, \citenamefont {Shchadilova},\ and\ \citenamefont {Demler}}]{PolaronDynamics4}%
  \BibitemOpen
  \bibfield  {author} {\bibinfo {author} {\bibfnamefont {F.}~\bibnamefont {Grusdt}}, \bibinfo {author} {\bibfnamefont {K.}~\bibnamefont {Seetharam}}, \bibinfo {author} {\bibfnamefont {Y.}~\bibnamefont {Shchadilova}}, \ and\ \bibinfo {author} {\bibfnamefont {E.}~\bibnamefont {Demler}},\ }\href {\doibase 10.1103/PhysRevA.97.033612} {\bibfield  {journal} {\bibinfo  {journal} {Phys. Rev. A}\ }\textbf {\bibinfo {volume} {97}},\ \bibinfo {pages} {033612} (\bibinfo {year} {2018})}\BibitemShut {NoStop}%
\bibitem [{\citenamefont {Cabrera}\ \emph {et~al.}(2018)\citenamefont {Cabrera}, \citenamefont {Tanzi}, \citenamefont {Sanz}, \citenamefont {Naylor}, \citenamefont {Thomas}, \citenamefont {Cheiney},\ and\ \citenamefont {Tarruell}}]{CabreraTarruellDropExp}%
  \BibitemOpen
  \bibfield  {author} {\bibinfo {author} {\bibfnamefont {C.~R.}\ \bibnamefont {Cabrera}}, \bibinfo {author} {\bibfnamefont {L.}~\bibnamefont {Tanzi}}, \bibinfo {author} {\bibfnamefont {J.}~\bibnamefont {Sanz}}, \bibinfo {author} {\bibfnamefont {B.}~\bibnamefont {Naylor}}, \bibinfo {author} {\bibfnamefont {P.}~\bibnamefont {Thomas}}, \bibinfo {author} {\bibfnamefont {P.}~\bibnamefont {Cheiney}}, \ and\ \bibinfo {author} {\bibfnamefont {L.}~\bibnamefont {Tarruell}},\ }\href {\doibase 10.1126/science.aao5686} {\bibfield  {journal} {\bibinfo  {journal} {Science}\ }\textbf {\bibinfo {volume} {359}},\ \bibinfo {pages} {301} (\bibinfo {year} {2018})}\BibitemShut {NoStop}%
\bibitem [{\citenamefont {Cheiney}\ \emph {et~al.}(2018)\citenamefont {Cheiney}, \citenamefont {Cabrera}, \citenamefont {Sanz}, \citenamefont {Naylor}, \citenamefont {Tanzi},\ and\ \citenamefont {Tarruell}}]{CheineyTarruellDropExp}%
  \BibitemOpen
  \bibfield  {author} {\bibinfo {author} {\bibfnamefont {P.}~\bibnamefont {Cheiney}}, \bibinfo {author} {\bibfnamefont {C.~R.}\ \bibnamefont {Cabrera}}, \bibinfo {author} {\bibfnamefont {J.}~\bibnamefont {Sanz}}, \bibinfo {author} {\bibfnamefont {B.}~\bibnamefont {Naylor}}, \bibinfo {author} {\bibfnamefont {L.}~\bibnamefont {Tanzi}}, \ and\ \bibinfo {author} {\bibfnamefont {L.}~\bibnamefont {Tarruell}},\ }\href {\doibase 10.1103/PhysRevLett.120.135301} {\bibfield  {journal} {\bibinfo  {journal} {Phys. Rev. Lett.}\ }\textbf {\bibinfo {volume} {120}},\ \bibinfo {pages} {135301} (\bibinfo {year} {2018})}\BibitemShut {NoStop}%
\bibitem [{\citenamefont {D'Errico}\ \emph {et~al.}(2019)\citenamefont {D'Errico}, \citenamefont {Burchianti}, \citenamefont {Prevedelli}, \citenamefont {Salasnich}, \citenamefont {Ancilotto}, \citenamefont {Modugno}, \citenamefont {Minardi},\ and\ \citenamefont {Fort}}]{FortHeteroExp}%
  \BibitemOpen
  \bibfield  {author} {\bibinfo {author} {\bibfnamefont {C.}~\bibnamefont {D'Errico}}, \bibinfo {author} {\bibfnamefont {A.}~\bibnamefont {Burchianti}}, \bibinfo {author} {\bibfnamefont {M.}~\bibnamefont {Prevedelli}}, \bibinfo {author} {\bibfnamefont {L.}~\bibnamefont {Salasnich}}, \bibinfo {author} {\bibfnamefont {F.}~\bibnamefont {Ancilotto}}, \bibinfo {author} {\bibfnamefont {M.}~\bibnamefont {Modugno}}, \bibinfo {author} {\bibfnamefont {F.}~\bibnamefont {Minardi}}, \ and\ \bibinfo {author} {\bibfnamefont {C.}~\bibnamefont {Fort}},\ }\href {\doibase 10.1103/PhysRevResearch.1.033155} {\bibfield  {journal} {\bibinfo  {journal} {Phys. Rev. Research}\ }\textbf {\bibinfo {volume} {1}},\ \bibinfo {pages} {033155} (\bibinfo {year} {2019})}\BibitemShut {NoStop}%
\bibitem [{\citenamefont {B\"ottcher}\ \emph {et~al.}(2019)\citenamefont {B\"ottcher}, \citenamefont {Schmidt}, \citenamefont {Wenzel}, \citenamefont {Hertkorn}, \citenamefont {Guo}, \citenamefont {Langen},\ and\ \citenamefont {Pfau}}]{Bottcher2019SupersolidDrop}%
  \BibitemOpen
  \bibfield  {author} {\bibinfo {author} {\bibfnamefont {F.}~\bibnamefont {B\"ottcher}}, \bibinfo {author} {\bibfnamefont {J.-N.}\ \bibnamefont {Schmidt}}, \bibinfo {author} {\bibfnamefont {M.}~\bibnamefont {Wenzel}}, \bibinfo {author} {\bibfnamefont {J.}~\bibnamefont {Hertkorn}}, \bibinfo {author} {\bibfnamefont {M.}~\bibnamefont {Guo}}, \bibinfo {author} {\bibfnamefont {T.}~\bibnamefont {Langen}}, \ and\ \bibinfo {author} {\bibfnamefont {T.}~\bibnamefont {Pfau}},\ }\href {\doibase 10.1103/PhysRevX.9.011051} {\bibfield  {journal} {\bibinfo  {journal} {Phys. Rev. X}\ }\textbf {\bibinfo {volume} {9}},\ \bibinfo {pages} {011051} (\bibinfo {year} {2019})}\BibitemShut {NoStop}%
\bibitem [{\citenamefont {Chomaz}\ \emph {et~al.}(2022)\citenamefont {Chomaz}, \citenamefont {Ferrier-Barbut}, \citenamefont {Ferlaino}, \citenamefont {Laburthe-Tolra}, \citenamefont {Lev},\ and\ \citenamefont {Pfau}}]{Chomaz_2023}%
  \BibitemOpen
  \bibfield  {author} {\bibinfo {author} {\bibfnamefont {L.}~\bibnamefont {Chomaz}}, \bibinfo {author} {\bibfnamefont {I.}~\bibnamefont {Ferrier-Barbut}}, \bibinfo {author} {\bibfnamefont {F.}~\bibnamefont {Ferlaino}}, \bibinfo {author} {\bibfnamefont {B.}~\bibnamefont {Laburthe-Tolra}}, \bibinfo {author} {\bibfnamefont {B.~L.}\ \bibnamefont {Lev}}, \ and\ \bibinfo {author} {\bibfnamefont {T.}~\bibnamefont {Pfau}},\ }\href {\doibase 10.1088/1361-6633/aca814} {\bibfield  {journal} {\bibinfo  {journal} {Rev. Mod. Phys.}\ }\textbf {\bibinfo {volume} {86}},\ \bibinfo {pages} {026401} (\bibinfo {year} {2022})}\BibitemShut {NoStop}%
\bibitem [{\citenamefont {Bisset}\ \emph {et~al.}(2021)\citenamefont {Bisset}, \citenamefont {Ardila},\ and\ \citenamefont {Santos}}]{Bisset2021}%
  \BibitemOpen
  \bibfield  {author} {\bibinfo {author} {\bibfnamefont {R.~N.}\ \bibnamefont {Bisset}}, \bibinfo {author} {\bibfnamefont {L.~A.~P.}\ \bibnamefont {Ardila}}, \ and\ \bibinfo {author} {\bibfnamefont {L.}~\bibnamefont {Santos}},\ }\href {\doibase 10.1103/PhysRevLett.126.025301} {\bibfield  {journal} {\bibinfo  {journal} {Phys. Rev. Lett.}\ }\textbf {\bibinfo {volume} {126}},\ \bibinfo {pages} {025301} (\bibinfo {year} {2021})}\BibitemShut {NoStop}%
\bibitem [{\citenamefont {Smith}\ \emph {et~al.}(2021)\citenamefont {Smith}, \citenamefont {Baillie},\ and\ \citenamefont {Blakie}}]{Smith2021}%
  \BibitemOpen
  \bibfield  {author} {\bibinfo {author} {\bibfnamefont {J.~C.}\ \bibnamefont {Smith}}, \bibinfo {author} {\bibfnamefont {D.}~\bibnamefont {Baillie}}, \ and\ \bibinfo {author} {\bibfnamefont {P.~B.}\ \bibnamefont {Blakie}},\ }\href {\doibase 10.1103/PhysRevLett.126.025302} {\bibfield  {journal} {\bibinfo  {journal} {Phys. Rev. Lett.}\ }\textbf {\bibinfo {volume} {126}},\ \bibinfo {pages} {025302} (\bibinfo {year} {2021})}\BibitemShut {NoStop}%
\bibitem [{\citenamefont {Lee}\ \emph {et~al.}(1957)\citenamefont {Lee}, \citenamefont {Huang},\ and\ \citenamefont {Yang}}]{LeeHuangYang1957}%
  \BibitemOpen
  \bibfield  {author} {\bibinfo {author} {\bibfnamefont {T.~D.}\ \bibnamefont {Lee}}, \bibinfo {author} {\bibfnamefont {K.}~\bibnamefont {Huang}}, \ and\ \bibinfo {author} {\bibfnamefont {C.~N.}\ \bibnamefont {Yang}},\ }\href {\doibase 10.1103/PhysRev.106.1135} {\bibfield  {journal} {\bibinfo  {journal} {Phys. Rev.}\ }\textbf {\bibinfo {volume} {106}},\ \bibinfo {pages} {1135} (\bibinfo {year} {1957})}\BibitemShut {NoStop}%
\bibitem [{\citenamefont {Sekino}\ and\ \citenamefont {Nishida}(2018)}]{Nishida3body}%
  \BibitemOpen
  \bibfield  {author} {\bibinfo {author} {\bibfnamefont {Y.}~\bibnamefont {Sekino}}\ and\ \bibinfo {author} {\bibfnamefont {Y.}~\bibnamefont {Nishida}},\ }\href {\doibase 10.1103/PhysRevA.97.011602} {\bibfield  {journal} {\bibinfo  {journal} {Phys. Rev. A}\ }\textbf {\bibinfo {volume} {97}},\ \bibinfo {pages} {011602} (\bibinfo {year} {2018})}\BibitemShut {NoStop}%
\bibitem [{\citenamefont {Morera}\ \emph {et~al.}(2022)\citenamefont {Morera}, \citenamefont {Juli\'a-D\'{\i}az},\ and\ \citenamefont {Valiente}}]{Morera3body1D}%
  \BibitemOpen
  \bibfield  {author} {\bibinfo {author} {\bibfnamefont {I.}~\bibnamefont {Morera}}, \bibinfo {author} {\bibfnamefont {B.}~\bibnamefont {Juli\'a-D\'{\i}az}}, \ and\ \bibinfo {author} {\bibfnamefont {M.}~\bibnamefont {Valiente}},\ }\href {\doibase 10.1103/PhysRevResearch.4.L042024} {\bibfield  {journal} {\bibinfo  {journal} {Phys. Rev. Res.}\ }\textbf {\bibinfo {volume} {4}},\ \bibinfo {pages} {L042024} (\bibinfo {year} {2022})}\BibitemShut {NoStop}%
\bibitem [{\citenamefont {Cui}(2018)}]{CuiSpinOrbitBoseFermi}%
  \BibitemOpen
  \bibfield  {author} {\bibinfo {author} {\bibfnamefont {X.}~\bibnamefont {Cui}},\ }\href {\doibase 10.1103/PhysRevA.98.023630} {\bibfield  {journal} {\bibinfo  {journal} {Phys. Rev. A}\ }\textbf {\bibinfo {volume} {98}},\ \bibinfo {pages} {023630} (\bibinfo {year} {2018})}\BibitemShut {NoStop}%
\bibitem [{\citenamefont {Rakshit}\ \emph {et~al.}(2019)\citenamefont {Rakshit}, \citenamefont {Karpiuk}, \citenamefont {Brewczyk},\ and\ \citenamefont {Gajda}}]{GajdaBoseFermi}%
  \BibitemOpen
  \bibfield  {author} {\bibinfo {author} {\bibfnamefont {D.}~\bibnamefont {Rakshit}}, \bibinfo {author} {\bibfnamefont {T.}~\bibnamefont {Karpiuk}}, \bibinfo {author} {\bibfnamefont {M.}~\bibnamefont {Brewczyk}}, \ and\ \bibinfo {author} {\bibfnamefont {M.}~\bibnamefont {Gajda}},\ }\href {\doibase 10.21468/SciPostPhys.6.6.079} {\bibfield  {journal} {\bibinfo  {journal} {SciPost Phys.}\ }\textbf {\bibinfo {volume} {6}},\ \bibinfo {pages} {79} (\bibinfo {year} {2019})}\BibitemShut {NoStop}%
\bibitem [{\citenamefont {Wang}\ \emph {et~al.}(2020)\citenamefont {Wang}, \citenamefont {Pan}, \citenamefont {Cui},\ and\ \citenamefont {Yi}}]{Wang_2020BoseFermi}%
  \BibitemOpen
  \bibfield  {author} {\bibinfo {author} {\bibfnamefont {J.-B.}\ \bibnamefont {Wang}}, \bibinfo {author} {\bibfnamefont {J.-S.}\ \bibnamefont {Pan}}, \bibinfo {author} {\bibfnamefont {X.}~\bibnamefont {Cui}}, \ and\ \bibinfo {author} {\bibfnamefont {W.}~\bibnamefont {Yi}},\ }\href {\doibase 10.1088/0256-307x/37/7/076701} {\bibfield  {journal} {\bibinfo  {journal} {Chin. Phys. Lett.}\ }\textbf {\bibinfo {volume} {37}},\ \bibinfo {pages} {076701} (\bibinfo {year} {2020})}\BibitemShut {NoStop}%
\bibitem [{\citenamefont {Ancilotto}\ \emph {et~al.}(2018)\citenamefont {Ancilotto}, \citenamefont {Barranco}, \citenamefont {Guilleumas},\ and\ \citenamefont {Pi}}]{AncilottoLocalDensity}%
  \BibitemOpen
  \bibfield  {author} {\bibinfo {author} {\bibfnamefont {F.}~\bibnamefont {Ancilotto}}, \bibinfo {author} {\bibfnamefont {M.}~\bibnamefont {Barranco}}, \bibinfo {author} {\bibfnamefont {M.}~\bibnamefont {Guilleumas}}, \ and\ \bibinfo {author} {\bibfnamefont {M.}~\bibnamefont {Pi}},\ }\href {\doibase 10.1103/PhysRevA.98.053623} {\bibfield  {journal} {\bibinfo  {journal} {Phys. Rev. A}\ }\textbf {\bibinfo {volume} {98}},\ \bibinfo {pages} {053623} (\bibinfo {year} {2018})}\BibitemShut {NoStop}%
\bibitem [{\citenamefont {Ferioli}\ \emph {et~al.}(2019)\citenamefont {Ferioli}, \citenamefont {Semeghini}, \citenamefont {Masi}, \citenamefont {Giusti}, \citenamefont {Modugno}, \citenamefont {Inguscio}, \citenamefont {Gallem\'{\i}}, \citenamefont {Recati},\ and\ \citenamefont {Fattori}}]{FattoriCollisions}%
  \BibitemOpen
  \bibfield  {author} {\bibinfo {author} {\bibfnamefont {G.}~\bibnamefont {Ferioli}}, \bibinfo {author} {\bibfnamefont {G.}~\bibnamefont {Semeghini}}, \bibinfo {author} {\bibfnamefont {L.}~\bibnamefont {Masi}}, \bibinfo {author} {\bibfnamefont {G.}~\bibnamefont {Giusti}}, \bibinfo {author} {\bibfnamefont {G.}~\bibnamefont {Modugno}}, \bibinfo {author} {\bibfnamefont {M.}~\bibnamefont {Inguscio}}, \bibinfo {author} {\bibfnamefont {A.}~\bibnamefont {Gallem\'{\i}}}, \bibinfo {author} {\bibfnamefont {A.}~\bibnamefont {Recati}}, \ and\ \bibinfo {author} {\bibfnamefont {M.}~\bibnamefont {Fattori}},\ }\href {\doibase 10.1103/PhysRevLett.122.090401} {\bibfield  {journal} {\bibinfo  {journal} {Phys. Rev. Lett.}\ }\textbf {\bibinfo {volume} {122}},\ \bibinfo {pages} {090401} (\bibinfo {year} {2019})}\BibitemShut {NoStop}%
\bibitem [{\citenamefont {Fort}\ and\ \citenamefont {Modugno}(2021)}]{FortModugnoSelfEvaporation}%
  \BibitemOpen
  \bibfield  {author} {\bibinfo {author} {\bibfnamefont {C.}~\bibnamefont {Fort}}\ and\ \bibinfo {author} {\bibfnamefont {M.}~\bibnamefont {Modugno}},\ }\href {\doibase 10.3390/app11020866} {\bibfield  {journal} {\bibinfo  {journal} {Appl. Sci.}\ }\textbf {\bibinfo {volume} {11(2)}},\ \bibinfo {pages} {866} (\bibinfo {year} {2021})}\BibitemShut {NoStop}%
\bibitem [{\citenamefont {Astrakharchik}\ and\ \citenamefont {Malomed}(2018)}]{AstrakharchikMalomed1DDynamics}%
  \BibitemOpen
  \bibfield  {author} {\bibinfo {author} {\bibfnamefont {G.~E.}\ \bibnamefont {Astrakharchik}}\ and\ \bibinfo {author} {\bibfnamefont {B.~A.}\ \bibnamefont {Malomed}},\ }\href {\doibase 10.1103/PhysRevA.98.013631} {\bibfield  {journal} {\bibinfo  {journal} {Phys. Rev. A}\ }\textbf {\bibinfo {volume} {98}},\ \bibinfo {pages} {013631} (\bibinfo {year} {2018})}\BibitemShut {NoStop}%
\bibitem [{\citenamefont {Semeghini}\ \emph {et~al.}(2018)\citenamefont {Semeghini}, \citenamefont {Ferioli}, \citenamefont {Masi}, \citenamefont {Mazzinghi}, \citenamefont {Wolswijk}, \citenamefont {Minardi}, \citenamefont {Modugno}, \citenamefont {Modugno}, \citenamefont {Inguscio},\ and\ \citenamefont {Fattori}}]{SemeghiniFattoriDropExp}%
  \BibitemOpen
  \bibfield  {author} {\bibinfo {author} {\bibfnamefont {G.}~\bibnamefont {Semeghini}}, \bibinfo {author} {\bibfnamefont {G.}~\bibnamefont {Ferioli}}, \bibinfo {author} {\bibfnamefont {L.}~\bibnamefont {Masi}}, \bibinfo {author} {\bibfnamefont {C.}~\bibnamefont {Mazzinghi}}, \bibinfo {author} {\bibfnamefont {L.}~\bibnamefont {Wolswijk}}, \bibinfo {author} {\bibfnamefont {F.}~\bibnamefont {Minardi}}, \bibinfo {author} {\bibfnamefont {M.}~\bibnamefont {Modugno}}, \bibinfo {author} {\bibfnamefont {G.}~\bibnamefont {Modugno}}, \bibinfo {author} {\bibfnamefont {M.}~\bibnamefont {Inguscio}}, \ and\ \bibinfo {author} {\bibfnamefont {M.}~\bibnamefont {Fattori}},\ }\href {\doibase 10.1103/PhysRevLett.120.235301} {\bibfield  {journal} {\bibinfo  {journal} {Phys. Rev. Lett.}\ }\textbf {\bibinfo {volume} {120}},\ \bibinfo {pages} {235301} (\bibinfo {year} {2018})}\BibitemShut {NoStop}%
\bibitem [{\citenamefont {Mistakidis}\ \emph {et~al.}(2021)\citenamefont {Mistakidis}, \citenamefont {Mithun}, \citenamefont {Kevrekidis}, \citenamefont {Sadeghpour},\ and\ \citenamefont {Schmelcher}}]{Mistakidis2021}%
  \BibitemOpen
  \bibfield  {author} {\bibinfo {author} {\bibfnamefont {S.~I.}\ \bibnamefont {Mistakidis}}, \bibinfo {author} {\bibfnamefont {T.}~\bibnamefont {Mithun}}, \bibinfo {author} {\bibfnamefont {P.~G.}\ \bibnamefont {Kevrekidis}}, \bibinfo {author} {\bibfnamefont {H.~R.}\ \bibnamefont {Sadeghpour}}, \ and\ \bibinfo {author} {\bibfnamefont {P.}~\bibnamefont {Schmelcher}},\ }\href {\doibase 10.1103/PhysRevResearch.3.043128} {\bibfield  {journal} {\bibinfo  {journal} {Phys. Rev. Research}\ }\textbf {\bibinfo {volume} {3}},\ \bibinfo {pages} {043128} (\bibinfo {year} {2021})}\BibitemShut {NoStop}%
\bibitem [{\citenamefont {Englezos}\ \emph {et~al.}(2023)\citenamefont {Englezos}, \citenamefont {Mistakidis},\ and\ \citenamefont {Schmelcher}}]{Englezos2023}%
  \BibitemOpen
  \bibfield  {author} {\bibinfo {author} {\bibfnamefont {I.~A.}\ \bibnamefont {Englezos}}, \bibinfo {author} {\bibfnamefont {S.~I.}\ \bibnamefont {Mistakidis}}, \ and\ \bibinfo {author} {\bibfnamefont {P.}~\bibnamefont {Schmelcher}},\ }\href {\doibase 10.1103/PhysRevA.107.023320} {\bibfield  {journal} {\bibinfo  {journal} {Phys. Rev. A}\ }\textbf {\bibinfo {volume} {107}},\ \bibinfo {pages} {023320} (\bibinfo {year} {2023})}\BibitemShut {NoStop}%
\bibitem [{\citenamefont {He}\ \emph {et~al.}(2023)\citenamefont {He}, \citenamefont {Li}, \citenamefont {Yi},\ and\ \citenamefont {Yu}}]{QuantumCrit2023}%
  \BibitemOpen
  \bibfield  {author} {\bibinfo {author} {\bibfnamefont {L.}~\bibnamefont {He}}, \bibinfo {author} {\bibfnamefont {H.}~\bibnamefont {Li}}, \bibinfo {author} {\bibfnamefont {W.}~\bibnamefont {Yi}}, \ and\ \bibinfo {author} {\bibfnamefont {Z.-Q.}\ \bibnamefont {Yu}},\ }\href {\doibase 10.1103/PhysRevLett.130.193001} {\bibfield  {journal} {\bibinfo  {journal} {Phys. Rev. Lett.}\ }\textbf {\bibinfo {volume} {130}},\ \bibinfo {pages} {193001} (\bibinfo {year} {2023})}\BibitemShut {NoStop}%
\bibitem [{\citenamefont {Flynn}\ \emph {et~al.}(2023{\natexlab{a}})\citenamefont {Flynn}, \citenamefont {Parisi}, \citenamefont {Billam},\ and\ \citenamefont {Parker}}]{FlynnPRR2023}%
  \BibitemOpen
  \bibfield  {author} {\bibinfo {author} {\bibfnamefont {T.~A.}\ \bibnamefont {Flynn}}, \bibinfo {author} {\bibfnamefont {L.}~\bibnamefont {Parisi}}, \bibinfo {author} {\bibfnamefont {T.~P.}\ \bibnamefont {Billam}}, \ and\ \bibinfo {author} {\bibfnamefont {N.~G.}\ \bibnamefont {Parker}},\ }\href {\doibase 10.1103/PhysRevResearch.5.033167} {\bibfield  {journal} {\bibinfo  {journal} {Phys. Rev. Research}\ }\textbf {\bibinfo {volume} {5}},\ \bibinfo {pages} {033167} (\bibinfo {year} {2023}{\natexlab{a}})}\BibitemShut {NoStop}%
\bibitem [{\citenamefont {Flynn}\ \emph {et~al.}(2023{\natexlab{b}})\citenamefont {Flynn}, \citenamefont {Keepfer}, \citenamefont {Billam},\ and\ \citenamefont {Parker}}]{FlynnTrapped2023}%
  \BibitemOpen
  \bibfield  {author} {\bibinfo {author} {\bibfnamefont {T.~A.}\ \bibnamefont {Flynn}}, \bibinfo {author} {\bibfnamefont {N.}~\bibnamefont {Keepfer}}, \bibinfo {author} {\bibfnamefont {T.~P.}\ \bibnamefont {Billam}}, \ and\ \bibinfo {author} {\bibfnamefont {N.~G.}\ \bibnamefont {Parker}},\ }\href {https://doi.org/10.48550/arXiv.2309.04300} {\bibfield  {journal} {\bibinfo  {journal} {arXiv:2309.04300}\ } (\bibinfo {year} {2023}{\natexlab{b}})}\BibitemShut {NoStop}%
\bibitem [{\citenamefont {Vallès-Muns}\ \emph {et~al.}(2023)\citenamefont {Vallès-Muns}, \citenamefont {Morera}, \citenamefont {Astrakharchik},\ and\ \citenamefont {Juliá-Díaz}}]{latticeDrop2023}%
  \BibitemOpen
  \bibfield  {author} {\bibinfo {author} {\bibfnamefont {J.}~\bibnamefont {Vallès-Muns}}, \bibinfo {author} {\bibfnamefont {I.}~\bibnamefont {Morera}}, \bibinfo {author} {\bibfnamefont {G.~E.}\ \bibnamefont {Astrakharchik}}, \ and\ \bibinfo {author} {\bibfnamefont {B.}~\bibnamefont {Juliá-Díaz}},\ }\href {https://arxiv.org/abs/2306.12283} {\bibfield  {journal} {\bibinfo  {journal} {arXiv:2306.12283}\ } (\bibinfo {year} {2023})}\BibitemShut {NoStop}%
\bibitem [{\citenamefont {Tengstrand}\ and\ \citenamefont {Reimann}(2022)}]{tengstrand2022droplet}%
  \BibitemOpen
  \bibfield  {author} {\bibinfo {author} {\bibfnamefont {M.~N.}\ \bibnamefont {Tengstrand}}\ and\ \bibinfo {author} {\bibfnamefont {S.}~\bibnamefont {Reimann}},\ }\href@noop {} {\bibfield  {journal} {\bibinfo  {journal} {Phys. Rev. A}\ }\textbf {\bibinfo {volume} {105}},\ \bibinfo {pages} {033319} (\bibinfo {year} {2022})}\BibitemShut {NoStop}%
\bibitem [{\citenamefont {Petrov}\ and\ \citenamefont {Astrakharchik}(2016)}]{PetrovLowD}%
  \BibitemOpen
  \bibfield  {author} {\bibinfo {author} {\bibfnamefont {D.~S.}\ \bibnamefont {Petrov}}\ and\ \bibinfo {author} {\bibfnamefont {G.~E.}\ \bibnamefont {Astrakharchik}},\ }\href {\doibase 10.1103/PhysRevLett.117.100401} {\bibfield  {journal} {\bibinfo  {journal} {Phys. Rev. Lett.}\ }\textbf {\bibinfo {volume} {117}},\ \bibinfo {pages} {100401} (\bibinfo {year} {2016})}\BibitemShut {NoStop}%
\bibitem [{\citenamefont {Krönke}\ \emph {et~al.}(2013)\citenamefont {Krönke}, \citenamefont {Cao}, \citenamefont {Vendrell},\ and\ \citenamefont {Schmelcher}}]{Kronke_2013}%
  \BibitemOpen
  \bibfield  {author} {\bibinfo {author} {\bibfnamefont {S.}~\bibnamefont {Krönke}}, \bibinfo {author} {\bibfnamefont {L.}~\bibnamefont {Cao}}, \bibinfo {author} {\bibfnamefont {O.}~\bibnamefont {Vendrell}}, \ and\ \bibinfo {author} {\bibfnamefont {P.}~\bibnamefont {Schmelcher}},\ }\href {\doibase 10.1088/1367-2630/15/6/063018} {\bibfield  {journal} {\bibinfo  {journal} {New J. Phys.}\ }\textbf {\bibinfo {volume} {15}},\ \bibinfo {pages} {063018} (\bibinfo {year} {2013})}\BibitemShut {NoStop}%
\bibitem [{\citenamefont {Cao}\ \emph {et~al.}(2013)\citenamefont {Cao}, \citenamefont {Krönke}, \citenamefont {Vendrell},\ and\ \citenamefont {Schmelcher}}]{Cao2013}%
  \BibitemOpen
  \bibfield  {author} {\bibinfo {author} {\bibfnamefont {L.}~\bibnamefont {Cao}}, \bibinfo {author} {\bibfnamefont {S.}~\bibnamefont {Krönke}}, \bibinfo {author} {\bibfnamefont {O.}~\bibnamefont {Vendrell}}, \ and\ \bibinfo {author} {\bibfnamefont {P.}~\bibnamefont {Schmelcher}},\ }\href {\doibase 10.1063/1.4821350} {\bibfield  {journal} {\bibinfo  {journal} {J. Chem. Phys.}\ }\textbf {\bibinfo {volume} {139}},\ \bibinfo {pages} {134103} (\bibinfo {year} {2013})}\BibitemShut {NoStop}%
\bibitem [{\citenamefont {Cao}\ \emph {et~al.}(2017)\citenamefont {Cao}, \citenamefont {Bolsinger}, \citenamefont {Mistakidis}, \citenamefont {Koutentakis}, \citenamefont {Kr{\"o}nke}, \citenamefont {Schurer},\ and\ \citenamefont {Schmelcher}}]{cao2017unified}%
  \BibitemOpen
  \bibfield  {author} {\bibinfo {author} {\bibfnamefont {L.}~\bibnamefont {Cao}}, \bibinfo {author} {\bibfnamefont {V.}~\bibnamefont {Bolsinger}}, \bibinfo {author} {\bibfnamefont {S.~I.}\ \bibnamefont {Mistakidis}}, \bibinfo {author} {\bibfnamefont {G.~M.}\ \bibnamefont {Koutentakis}}, \bibinfo {author} {\bibfnamefont {S.}~\bibnamefont {Kr{\"o}nke}}, \bibinfo {author} {\bibfnamefont {J.~M.}\ \bibnamefont {Schurer}}, \ and\ \bibinfo {author} {\bibfnamefont {P.}~\bibnamefont {Schmelcher}},\ }\href@noop {} {\bibfield  {journal} {\bibinfo  {journal} {J. Chem. Phys.}\ }\textbf {\bibinfo {volume} {147}},\ \bibinfo {pages} {044106} (\bibinfo {year} {2017})}\BibitemShut {NoStop}%
\bibitem [{\citenamefont {Mistakidis}\ \emph {et~al.}(2023)\citenamefont {Mistakidis}, \citenamefont {Volosniev}, \citenamefont {Barfknecht}, \citenamefont {Fogarty}, \citenamefont {Busch}, \citenamefont {Foerster}, \citenamefont {Schmelcher},\ and\ \citenamefont {Zinner}}]{mistakidis2023few}%
  \BibitemOpen
  \bibfield  {author} {\bibinfo {author} {\bibfnamefont {S.~I.}\ \bibnamefont {Mistakidis}}, \bibinfo {author} {\bibfnamefont {A.~G.}\ \bibnamefont {Volosniev}}, \bibinfo {author} {\bibfnamefont {R.~E.}\ \bibnamefont {Barfknecht}}, \bibinfo {author} {\bibfnamefont {T.}~\bibnamefont {Fogarty}}, \bibinfo {author} {\bibfnamefont {T.}~\bibnamefont {Busch}}, \bibinfo {author} {\bibfnamefont {A.}~\bibnamefont {Foerster}}, \bibinfo {author} {\bibfnamefont {P.}~\bibnamefont {Schmelcher}}, \ and\ \bibinfo {author} {\bibfnamefont {N.~T.}\ \bibnamefont {Zinner}},\ }\href@noop {} {\bibfield  {journal} {\bibinfo  {journal} {Phys. Rep.}\ }\textbf {\bibinfo {volume} {1042}},\ \bibinfo {pages} {1} (\bibinfo {year} {2023})}\BibitemShut {NoStop}%
\bibitem [{\citenamefont {Lode}\ \emph {et~al.}(2020)\citenamefont {Lode}, \citenamefont {L{\'e}v{\^e}que}, \citenamefont {Madsen}, \citenamefont {Streltsov},\ and\ \citenamefont {Alon}}]{lode2020colloquium}%
  \BibitemOpen
  \bibfield  {author} {\bibinfo {author} {\bibfnamefont {A.~U.~J.}\ \bibnamefont {Lode}}, \bibinfo {author} {\bibfnamefont {C.}~\bibnamefont {L{\'e}v{\^e}que}}, \bibinfo {author} {\bibfnamefont {L.~B.}\ \bibnamefont {Madsen}}, \bibinfo {author} {\bibfnamefont {A.~I.}\ \bibnamefont {Streltsov}}, \ and\ \bibinfo {author} {\bibfnamefont {O.~E.}\ \bibnamefont {Alon}},\ }\href@noop {} {\bibfield  {journal} {\bibinfo  {journal} {Rev. Mod. Phys.}\ }\textbf {\bibinfo {volume} {92}},\ \bibinfo {pages} {011001} (\bibinfo {year} {2020})}\BibitemShut {NoStop}%
\bibitem [{\citenamefont {Bakkali-Hassani}\ \emph {et~al.}(2021)\citenamefont {Bakkali-Hassani}, \citenamefont {Maury}, \citenamefont {Zou}, \citenamefont {Le~Cerf}, \citenamefont {Saint-Jalm}, \citenamefont {Castilho}, \citenamefont {Nascimbene}, \citenamefont {Dalibard},\ and\ \citenamefont {Beugnon}}]{BakkaliTowns2021}%
  \BibitemOpen
  \bibfield  {author} {\bibinfo {author} {\bibfnamefont {B.}~\bibnamefont {Bakkali-Hassani}}, \bibinfo {author} {\bibfnamefont {C.}~\bibnamefont {Maury}}, \bibinfo {author} {\bibfnamefont {Y.-Q.}\ \bibnamefont {Zou}}, \bibinfo {author} {\bibfnamefont {E.}~\bibnamefont {Le~Cerf}}, \bibinfo {author} {\bibfnamefont {R.}~\bibnamefont {Saint-Jalm}}, \bibinfo {author} {\bibfnamefont {P.~C.~M.}\ \bibnamefont {Castilho}}, \bibinfo {author} {\bibfnamefont {S.}~\bibnamefont {Nascimbene}}, \bibinfo {author} {\bibfnamefont {J.}~\bibnamefont {Dalibard}}, \ and\ \bibinfo {author} {\bibfnamefont {J.}~\bibnamefont {Beugnon}},\ }\href {\doibase 10.1103/PhysRevLett.127.023603} {\bibfield  {journal} {\bibinfo  {journal} {Phys. Rev. Lett.}\ }\textbf {\bibinfo {volume} {127}},\ \bibinfo {pages} {023603} (\bibinfo {year} {2021})}\BibitemShut {NoStop}%
\bibitem [{\citenamefont {Olshanii}(1998)}]{olshanii1998atomic}%
  \BibitemOpen
  \bibfield  {author} {\bibinfo {author} {\bibfnamefont {M.}~\bibnamefont {Olshanii}},\ }\href {\doibase 10.1103/PhysRevLett.81.938} {\bibfield  {journal} {\bibinfo  {journal} {Phys. Rev. Lett.}\ }\textbf {\bibinfo {volume} {81}},\ \bibinfo {pages} {938} (\bibinfo {year} {1998})}\BibitemShut {NoStop}%
\bibitem [{\citenamefont {Chin}\ \emph {et~al.}(2010)\citenamefont {Chin}, \citenamefont {Grimm}, \citenamefont {Julienne},\ and\ \citenamefont {Tiesinga}}]{chin2010feshbach}%
  \BibitemOpen
  \bibfield  {author} {\bibinfo {author} {\bibfnamefont {C.}~\bibnamefont {Chin}}, \bibinfo {author} {\bibfnamefont {R.}~\bibnamefont {Grimm}}, \bibinfo {author} {\bibfnamefont {P.}~\bibnamefont {Julienne}}, \ and\ \bibinfo {author} {\bibfnamefont {E.}~\bibnamefont {Tiesinga}},\ }\href@noop {} {\bibfield  {journal} {\bibinfo  {journal} {Rev. Mod. Phys.}\ }\textbf {\bibinfo {volume} {82}},\ \bibinfo {pages} {1225} (\bibinfo {year} {2010})}\BibitemShut {NoStop}%
\bibitem [{\citenamefont {K{\"o}hler}\ \emph {et~al.}(2006)\citenamefont {K{\"o}hler}, \citenamefont {G{\'o}ral},\ and\ \citenamefont {Julienne}}]{kohler2006production}%
  \BibitemOpen
  \bibfield  {author} {\bibinfo {author} {\bibfnamefont {T.}~\bibnamefont {K{\"o}hler}}, \bibinfo {author} {\bibfnamefont {K.}~\bibnamefont {G{\'o}ral}}, \ and\ \bibinfo {author} {\bibfnamefont {P.~S.}\ \bibnamefont {Julienne}},\ }\href@noop {} {\bibfield  {journal} {\bibinfo  {journal} {Rev. Mod. Phys.}\ }\textbf {\bibinfo {volume} {78}},\ \bibinfo {pages} {1311} (\bibinfo {year} {2006})}\BibitemShut {NoStop}%
\bibitem [{\citenamefont {G\"orlitz}\ \emph {et~al.}(2001)\citenamefont {G\"orlitz}, \citenamefont {Vogels}, \citenamefont {Leanhardt}, \citenamefont {Raman}, \citenamefont {Gustavson}, \citenamefont {Abo-Shaeer}, \citenamefont {Chikkatur}, \citenamefont {Gupta}, \citenamefont {Inouye}, \citenamefont {Rosenband},\ and\ \citenamefont {Ketterle}}]{Ketterle2001LowDexp}%
  \BibitemOpen
  \bibfield  {author} {\bibinfo {author} {\bibfnamefont {A.}~\bibnamefont {G\"orlitz}}, \bibinfo {author} {\bibfnamefont {J.~M.}\ \bibnamefont {Vogels}}, \bibinfo {author} {\bibfnamefont {A.~E.}\ \bibnamefont {Leanhardt}}, \bibinfo {author} {\bibfnamefont {C.}~\bibnamefont {Raman}}, \bibinfo {author} {\bibfnamefont {T.~L.}\ \bibnamefont {Gustavson}}, \bibinfo {author} {\bibfnamefont {J.~R.}\ \bibnamefont {Abo-Shaeer}}, \bibinfo {author} {\bibfnamefont {A.~P.}\ \bibnamefont {Chikkatur}}, \bibinfo {author} {\bibfnamefont {S.}~\bibnamefont {Gupta}}, \bibinfo {author} {\bibfnamefont {S.}~\bibnamefont {Inouye}}, \bibinfo {author} {\bibfnamefont {T.}~\bibnamefont {Rosenband}}, \ and\ \bibinfo {author} {\bibfnamefont {W.}~\bibnamefont {Ketterle}},\ }\href {\doibase 10.1103/PhysRevLett.87.130402} {\bibfield  {journal} {\bibinfo  {journal} {Phys. Rev. Lett.}\ }\textbf {\bibinfo {volume} {87}},\ \bibinfo {pages} {130402} (\bibinfo {year} {2001})}\BibitemShut {NoStop}%
\bibitem [{\citenamefont {Romero-Ros}\ \emph {et~al.}(2023)\citenamefont {Romero-Ros}, \citenamefont {Katsimiga}, \citenamefont {Mistakidis}, \citenamefont {Mossman}, \citenamefont {Biondini}, \citenamefont {Schmelcher}, \citenamefont {Engels},\ and\ \citenamefont {Kevrekidis}}]{romero2023experimental}%
  \BibitemOpen
  \bibfield  {author} {\bibinfo {author} {\bibfnamefont {A.}~\bibnamefont {Romero-Ros}}, \bibinfo {author} {\bibfnamefont {G.~C.}\ \bibnamefont {Katsimiga}}, \bibinfo {author} {\bibfnamefont {S.~I.}\ \bibnamefont {Mistakidis}}, \bibinfo {author} {\bibfnamefont {S.}~\bibnamefont {Mossman}}, \bibinfo {author} {\bibfnamefont {G.}~\bibnamefont {Biondini}}, \bibinfo {author} {\bibfnamefont {P.}~\bibnamefont {Schmelcher}}, \bibinfo {author} {\bibfnamefont {P.}~\bibnamefont {Engels}}, \ and\ \bibinfo {author} {\bibfnamefont {P.~G.}\ \bibnamefont {Kevrekidis}},\ }\href@noop {} {\bibfield  {journal} {\bibinfo  {journal} {arXiv preprint arXiv:2304.05951}\ } (\bibinfo {year} {2023})}\BibitemShut {NoStop}%
\bibitem [{\citenamefont {Mithun}\ \emph {et~al.}(2020)\citenamefont {Mithun}, \citenamefont {Maluckov}, \citenamefont {Kasamatsu}, \citenamefont {Malomed},\ and\ \citenamefont {Khare}}]{MithunMI}%
  \BibitemOpen
  \bibfield  {author} {\bibinfo {author} {\bibfnamefont {T.}~\bibnamefont {Mithun}}, \bibinfo {author} {\bibfnamefont {A.}~\bibnamefont {Maluckov}}, \bibinfo {author} {\bibfnamefont {K.}~\bibnamefont {Kasamatsu}}, \bibinfo {author} {\bibfnamefont {B.~A.}\ \bibnamefont {Malomed}}, \ and\ \bibinfo {author} {\bibfnamefont {A.}~\bibnamefont {Khare}},\ }\href {\doibase 10.3390/sym12010174} {\bibfield  {journal} {\bibinfo  {journal} {Symmetry}\ }\textbf {\bibinfo {volume} {12}},\ \bibinfo {pages} {32201} (\bibinfo {year} {2020})}\BibitemShut {NoStop}%
\bibitem [{\citenamefont {Parisi}\ \emph {et~al.}(2019)\citenamefont {Parisi}, \citenamefont {Astrakharchik},\ and\ \citenamefont {Giorgini}}]{ParisiMonteCarlo2019}%
  \BibitemOpen
  \bibfield  {author} {\bibinfo {author} {\bibfnamefont {L.}~\bibnamefont {Parisi}}, \bibinfo {author} {\bibfnamefont {G.~E.}\ \bibnamefont {Astrakharchik}}, \ and\ \bibinfo {author} {\bibfnamefont {S.}~\bibnamefont {Giorgini}},\ }\href {\doibase 10.1103/PhysRevLett.122.105302} {\bibfield  {journal} {\bibinfo  {journal} {Phys. Rev. Lett.}\ }\textbf {\bibinfo {volume} {122}},\ \bibinfo {pages} {105302} (\bibinfo {year} {2019})}\BibitemShut {NoStop}%
\bibitem [{\citenamefont {Parisi}\ and\ \citenamefont {Giorgini}(2020)}]{ParisiGiorginiMonteCarlo}%
  \BibitemOpen
  \bibfield  {author} {\bibinfo {author} {\bibfnamefont {L.}~\bibnamefont {Parisi}}\ and\ \bibinfo {author} {\bibfnamefont {S.}~\bibnamefont {Giorgini}},\ }\href {\doibase 10.1103/PhysRevA.102.023318} {\bibfield  {journal} {\bibinfo  {journal} {Phys. Rev. A}\ }\textbf {\bibinfo {volume} {102}},\ \bibinfo {pages} {023318} (\bibinfo {year} {2020})}\BibitemShut {NoStop}%
\bibitem [{\citenamefont {Horodecki}\ \emph {et~al.}(2009)\citenamefont {Horodecki}, \citenamefont {Horodecki}, \citenamefont {Horodecki},\ and\ \citenamefont {Horodecki}}]{horodecki2009quantum}%
  \BibitemOpen
  \bibfield  {author} {\bibinfo {author} {\bibfnamefont {R.}~\bibnamefont {Horodecki}}, \bibinfo {author} {\bibfnamefont {P.}~\bibnamefont {Horodecki}}, \bibinfo {author} {\bibfnamefont {M.}~\bibnamefont {Horodecki}}, \ and\ \bibinfo {author} {\bibfnamefont {K.}~\bibnamefont {Horodecki}},\ }\href@noop {} {\bibfield  {journal} {\bibinfo  {journal} {Rev. Mod. Phys.}\ }\textbf {\bibinfo {volume} {81}},\ \bibinfo {pages} {865} (\bibinfo {year} {2009})}\BibitemShut {NoStop}%
\bibitem [{\citenamefont {Mistakidis}\ \emph {et~al.}(2018)\citenamefont {Mistakidis}, \citenamefont {Katsimiga}, \citenamefont {Kevrekidis},\ and\ \citenamefont {Schmelcher}}]{mistakidis2018correlation}%
  \BibitemOpen
  \bibfield  {author} {\bibinfo {author} {\bibfnamefont {S.~I.}\ \bibnamefont {Mistakidis}}, \bibinfo {author} {\bibfnamefont {G.~C.}\ \bibnamefont {Katsimiga}}, \bibinfo {author} {\bibfnamefont {P.~G.}\ \bibnamefont {Kevrekidis}}, \ and\ \bibinfo {author} {\bibfnamefont {P.}~\bibnamefont {Schmelcher}},\ }\href@noop {} {\bibfield  {journal} {\bibinfo  {journal} {New J. Phys.}\ }\textbf {\bibinfo {volume} {20}},\ \bibinfo {pages} {043052} (\bibinfo {year} {2018})}\BibitemShut {NoStop}%
\bibitem [{\citenamefont {Frenkel}(1934)}]{frenkel1934wave}%
  \BibitemOpen
  \bibfield  {author} {\bibinfo {author} {\bibfnamefont {J.}~\bibnamefont {Frenkel}},\ }\href@noop {} {\enquote {\bibinfo {title} {Wave mechanics; elementary theory},}\ } (\bibinfo {year} {1934})\BibitemShut {NoStop}%
\bibitem [{\citenamefont {Pethick}\ and\ \citenamefont {Smith}(2008)}]{pethick2008bose}%
  \BibitemOpen
  \bibfield  {author} {\bibinfo {author} {\bibfnamefont {C.~J.}\ \bibnamefont {Pethick}}\ and\ \bibinfo {author} {\bibfnamefont {H.}~\bibnamefont {Smith}},\ }\href {\doibase 10.1017/CBO9780511802850} {\emph {\bibinfo {title} {Bose--Einstein condensation in dilute gases}}},\ \bibinfo {edition} {2nd}\ ed.\ (\bibinfo  {publisher} {Cambridge University Press},\ \bibinfo {year} {2008})\BibitemShut {NoStop}%
\bibitem [{\citenamefont {Pitaevskii}\ and\ \citenamefont {Stringari}(2016)}]{Stringari2016BEC}%
  \BibitemOpen
  \bibfield  {author} {\bibinfo {author} {\bibfnamefont {L.}~\bibnamefont {Pitaevskii}}\ and\ \bibinfo {author} {\bibfnamefont {S.}~\bibnamefont {Stringari}},\ }\href {https://global.oup.com/academic/product/bose-einstein-condensation-and-superfluidity-9780198758884?cc=de&lang=en&} {\emph {\bibinfo {title} {Bose–-Einstein Condensation and Superfluidity}}}\ (\bibinfo  {publisher} {Oxford University Press},\ \bibinfo {year} {2016})\BibitemShut {NoStop}%
\bibitem [{\citenamefont {Tylutki}\ \emph {et~al.}(2020)\citenamefont {Tylutki}, \citenamefont {Astrakharchik}, \citenamefont {Malomed},\ and\ \citenamefont {Petrov}}]{Collective1D}%
  \BibitemOpen
  \bibfield  {author} {\bibinfo {author} {\bibfnamefont {M.}~\bibnamefont {Tylutki}}, \bibinfo {author} {\bibfnamefont {G.~E.}\ \bibnamefont {Astrakharchik}}, \bibinfo {author} {\bibfnamefont {B.~A.}\ \bibnamefont {Malomed}}, \ and\ \bibinfo {author} {\bibfnamefont {D.~S.}\ \bibnamefont {Petrov}},\ }\href {\doibase 10.1103/PhysRevA.101.051601} {\bibfield  {journal} {\bibinfo  {journal} {Phys. Rev. A}\ }\textbf {\bibinfo {volume} {101}},\ \bibinfo {pages} {051601} (\bibinfo {year} {2020})}\BibitemShut {NoStop}%
\bibitem [{\citenamefont {P\'erez-Garc\'{\i}a}\ and\ \citenamefont {Beitia}(2005)}]{SymbioticSolitons}%
  \BibitemOpen
  \bibfield  {author} {\bibinfo {author} {\bibfnamefont {V.~M.}\ \bibnamefont {P\'erez-Garc\'{\i}a}}\ and\ \bibinfo {author} {\bibfnamefont {J.~B.}\ \bibnamefont {Beitia}},\ }\href {\doibase 10.1103/PhysRevA.72.033620} {\bibfield  {journal} {\bibinfo  {journal} {Phys. Rev. A}\ }\textbf {\bibinfo {volume} {72}},\ \bibinfo {pages} {033620} (\bibinfo {year} {2005})}\BibitemShut {NoStop}%
\bibitem [{\citenamefont {Abdullaev}\ and\ \citenamefont {Garnier}(2008)}]{Abdullaev2008}%
  \BibitemOpen
  \bibfield  {author} {\bibinfo {author} {\bibfnamefont {F.~K.}\ \bibnamefont {Abdullaev}}\ and\ \bibinfo {author} {\bibfnamefont {J.}~\bibnamefont {Garnier}},\ }\enquote {\bibinfo {title} {Bright solitons in bose-einstein condensates: Theory},}\ in\ \href {\doibase 10.1007/978-3-540-73591-5_2} {\emph {\bibinfo {booktitle} {Emergent Nonlinear Phenomena in Bose-Einstein Condensates: Theory and Experiment}}},\ \bibinfo {editor} {edited by\ \bibinfo {editor} {\bibfnamefont {P.~G.}\ \bibnamefont {Kevrekidis}}, \bibinfo {editor} {\bibfnamefont {D.~J.}\ \bibnamefont {Frantzeskakis}}, \ and\ \bibinfo {editor} {\bibfnamefont {R.}~\bibnamefont {Carretero-Gonz{\'a}lez}}}\ (\bibinfo  {publisher} {Springer Berlin Heidelberg},\ \bibinfo {address} {Berlin, Heidelberg},\ \bibinfo {year} {2008})\ pp.\ \bibinfo {pages} {25--43}\BibitemShut {NoStop}%
\bibitem [{\citenamefont {Sakmann}\ \emph {et~al.}(2008)\citenamefont {Sakmann}, \citenamefont {Streltsov}, \citenamefont {Alon},\ and\ \citenamefont {Cederbaum}}]{Sakmann2008Coher}%
  \BibitemOpen
  \bibfield  {author} {\bibinfo {author} {\bibfnamefont {K.}~\bibnamefont {Sakmann}}, \bibinfo {author} {\bibfnamefont {A.~I.}\ \bibnamefont {Streltsov}}, \bibinfo {author} {\bibfnamefont {O.~E.}\ \bibnamefont {Alon}}, \ and\ \bibinfo {author} {\bibfnamefont {L.~S.}\ \bibnamefont {Cederbaum}},\ }\href {\doibase 10.1103/PhysRevA.78.023615} {\bibfield  {journal} {\bibinfo  {journal} {Phys. Rev. A}\ }\textbf {\bibinfo {volume} {78}},\ \bibinfo {pages} {023615} (\bibinfo {year} {2008})}\BibitemShut {NoStop}%
\bibitem [{\citenamefont {Naraschewski}\ and\ \citenamefont {Glauber}(1999)}]{Naraschewski1999Coher}%
  \BibitemOpen
  \bibfield  {author} {\bibinfo {author} {\bibfnamefont {M.}~\bibnamefont {Naraschewski}}\ and\ \bibinfo {author} {\bibfnamefont {R.~J.}\ \bibnamefont {Glauber}},\ }\href {\doibase 10.1103/PhysRevA.59.4595} {\bibfield  {journal} {\bibinfo  {journal} {Phys. Rev. A}\ }\textbf {\bibinfo {volume} {59}},\ \bibinfo {pages} {4595} (\bibinfo {year} {1999})}\BibitemShut {NoStop}%
\bibitem [{\citenamefont {Mishmash}\ and\ \citenamefont {Carr}(2009)}]{MishmashPRL}%
  \BibitemOpen
  \bibfield  {author} {\bibinfo {author} {\bibfnamefont {R.~V.}\ \bibnamefont {Mishmash}}\ and\ \bibinfo {author} {\bibfnamefont {L.~D.}\ \bibnamefont {Carr}},\ }\href {\doibase 10.1103/PhysRevLett.103.140403} {\bibfield  {journal} {\bibinfo  {journal} {Phys. Rev. Lett.}\ }\textbf {\bibinfo {volume} {103}},\ \bibinfo {pages} {140403} (\bibinfo {year} {2009})}\BibitemShut {NoStop}%
\bibitem [{\citenamefont {Kr\"onke}\ and\ \citenamefont {Schmelcher}(2015{\natexlab{a}})}]{KronkePRAMBsolitons}%
  \BibitemOpen
  \bibfield  {author} {\bibinfo {author} {\bibfnamefont {S.}~\bibnamefont {Kr\"onke}}\ and\ \bibinfo {author} {\bibfnamefont {P.}~\bibnamefont {Schmelcher}},\ }\href {\doibase 10.1103/PhysRevA.91.053614} {\bibfield  {journal} {\bibinfo  {journal} {Phys. Rev. A}\ }\textbf {\bibinfo {volume} {91}},\ \bibinfo {pages} {053614} (\bibinfo {year} {2015}{\natexlab{a}})}\BibitemShut {NoStop}%
\bibitem [{\citenamefont {Kr\"onke}\ and\ \citenamefont {Schmelcher}(2015{\natexlab{b}})}]{KronkePRAtwoBodySoliton}%
  \BibitemOpen
  \bibfield  {author} {\bibinfo {author} {\bibfnamefont {S.}~\bibnamefont {Kr\"onke}}\ and\ \bibinfo {author} {\bibfnamefont {P.}~\bibnamefont {Schmelcher}},\ }\href {\doibase 10.1103/PhysRevA.92.023631} {\bibfield  {journal} {\bibinfo  {journal} {Phys. Rev. A}\ }\textbf {\bibinfo {volume} {92}},\ \bibinfo {pages} {023631} (\bibinfo {year} {2015}{\natexlab{b}})}\BibitemShut {NoStop}%
\bibitem [{\citenamefont {Katsimiga}\ \emph {et~al.}(2017)\citenamefont {Katsimiga}, \citenamefont {Mistakidis}, \citenamefont {Koutentakis}, \citenamefont {Kevrekidis},\ and\ \citenamefont {Schmelcher}}]{Katsimiga_2017}%
  \BibitemOpen
  \bibfield  {author} {\bibinfo {author} {\bibfnamefont {G.~C.}\ \bibnamefont {Katsimiga}}, \bibinfo {author} {\bibfnamefont {S.~I.}\ \bibnamefont {Mistakidis}}, \bibinfo {author} {\bibfnamefont {G.~M.}\ \bibnamefont {Koutentakis}}, \bibinfo {author} {\bibfnamefont {P.~G.}\ \bibnamefont {Kevrekidis}}, \ and\ \bibinfo {author} {\bibfnamefont {P.}~\bibnamefont {Schmelcher}},\ }\href {\doibase 10.1088/1367-2630/aa96f6} {\bibfield  {journal} {\bibinfo  {journal} {New J. Phys.}\ }\textbf {\bibinfo {volume} {19}},\ \bibinfo {pages} {123012} (\bibinfo {year} {2017})}\BibitemShut {NoStop}%
\bibitem [{\citenamefont {Katsimiga}\ \emph {et~al.}(2023)\citenamefont {Katsimiga}, \citenamefont {Mistakidis}, \citenamefont {Koutsokostas}, \citenamefont {Frantzeskakis}, \citenamefont {Carretero-Gonz{\'a}lez},\ and\ \citenamefont {Kevrekidis}}]{katsimiga2023solitary}%
  \BibitemOpen
  \bibfield  {author} {\bibinfo {author} {\bibfnamefont {G.~C.}\ \bibnamefont {Katsimiga}}, \bibinfo {author} {\bibfnamefont {S.~I.}\ \bibnamefont {Mistakidis}}, \bibinfo {author} {\bibfnamefont {G.~N.}\ \bibnamefont {Koutsokostas}}, \bibinfo {author} {\bibfnamefont {D.~J.}\ \bibnamefont {Frantzeskakis}}, \bibinfo {author} {\bibfnamefont {R.}~\bibnamefont {Carretero-Gonz{\'a}lez}}, \ and\ \bibinfo {author} {\bibfnamefont {P.~G.}\ \bibnamefont {Kevrekidis}},\ }\href@noop {} {\bibfield  {journal} {\bibinfo  {journal} {Phys. Rev. A}\ }\textbf {\bibinfo {volume} {107}},\ \bibinfo {pages} {063308} (\bibinfo {year} {2023})}\BibitemShut {NoStop}%
\bibitem [{\citenamefont {Mistakidis}\ \emph {et~al.}(2019{\natexlab{c}})\citenamefont {Mistakidis}, \citenamefont {Hilbig},\ and\ \citenamefont {Schmelcher}}]{Mistakidis2019Fermions}%
  \BibitemOpen
  \bibfield  {author} {\bibinfo {author} {\bibfnamefont {S.~I.}\ \bibnamefont {Mistakidis}}, \bibinfo {author} {\bibfnamefont {L.}~\bibnamefont {Hilbig}}, \ and\ \bibinfo {author} {\bibfnamefont {P.}~\bibnamefont {Schmelcher}},\ }\href {\doibase 10.1103/PhysRevA.100.023620} {\bibfield  {journal} {\bibinfo  {journal} {Phys. Rev. A}\ }\textbf {\bibinfo {volume} {100}},\ \bibinfo {pages} {023620} (\bibinfo {year} {2019}{\natexlab{c}})}\BibitemShut {NoStop}%
\bibitem [{\citenamefont {Karpiuk}\ \emph {et~al.}(2004)\citenamefont {Karpiuk}, \citenamefont {Brewczyk}, \citenamefont {Ospelkaus-Schwarzer}, \citenamefont {Bongs}, \citenamefont {Gajda},\ and\ \citenamefont {Rza\ifmmode \mbox{\c{}}\else \c{}\fi{}\ifmmode~\dot{z}\else \.{z}\fi{}ewski}}]{Karpiuk2004}%
  \BibitemOpen
  \bibfield  {author} {\bibinfo {author} {\bibfnamefont {T.}~\bibnamefont {Karpiuk}}, \bibinfo {author} {\bibfnamefont {M.}~\bibnamefont {Brewczyk}}, \bibinfo {author} {\bibfnamefont {S.}~\bibnamefont {Ospelkaus-Schwarzer}}, \bibinfo {author} {\bibfnamefont {K.}~\bibnamefont {Bongs}}, \bibinfo {author} {\bibfnamefont {M.}~\bibnamefont {Gajda}}, \ and\ \bibinfo {author} {\bibfnamefont {K.}~\bibnamefont {Rza\ifmmode \mbox{\c{}}\else \c{}\fi{}\ifmmode~\dot{z}\else \.{z}\fi{}ewski}},\ }\href {\doibase 10.1103/PhysRevLett.93.100401} {\bibfield  {journal} {\bibinfo  {journal} {Phys. Rev. Lett.}\ }\textbf {\bibinfo {volume} {93}},\ \bibinfo {pages} {100401} (\bibinfo {year} {2004})}\BibitemShut {NoStop}%
\bibitem [{\citenamefont {Hu}\ \emph {et~al.}(2020)\citenamefont {Hu}, \citenamefont {Wang},\ and\ \citenamefont {Liu}}]{Hui2020LowD}%
  \BibitemOpen
  \bibfield  {author} {\bibinfo {author} {\bibfnamefont {H.}~\bibnamefont {Hu}}, \bibinfo {author} {\bibfnamefont {J.}~\bibnamefont {Wang}}, \ and\ \bibinfo {author} {\bibfnamefont {X.-J.}\ \bibnamefont {Liu}},\ }\href {\doibase 10.1103/PhysRevA.102.043301} {\bibfield  {journal} {\bibinfo  {journal} {Phys. Rev. A}\ }\textbf {\bibinfo {volume} {102}},\ \bibinfo {pages} {043301} (\bibinfo {year} {2020})}\BibitemShut {NoStop}%
\bibitem [{\citenamefont {Gangwar}\ \emph {et~al.}(2023)\citenamefont {Gangwar}, \citenamefont {Ravisankar}, \citenamefont {Mistakidis}, \citenamefont {Muruganandam},\ and\ \citenamefont {Mishra}}]{gangwar2023spectrum}%
  \BibitemOpen
  \bibfield  {author} {\bibinfo {author} {\bibfnamefont {S.}~\bibnamefont {Gangwar}}, \bibinfo {author} {\bibfnamefont {R.}~\bibnamefont {Ravisankar}}, \bibinfo {author} {\bibfnamefont {S.~I.}\ \bibnamefont {Mistakidis}}, \bibinfo {author} {\bibfnamefont {P.}~\bibnamefont {Muruganandam}}, \ and\ \bibinfo {author} {\bibfnamefont {P.~K.}\ \bibnamefont {Mishra}},\ }\href@noop {} {\bibfield  {journal} {\bibinfo  {journal} {arXiv:2307.16742}\ } (\bibinfo {year} {2023})}\BibitemShut {NoStop}%
\bibitem [{\citenamefont {Chergui}\ \emph {et~al.}(2023)\citenamefont {Chergui}, \citenamefont {Bengtsson}, \citenamefont {Bjerlin}, \citenamefont {St{\"u}rmer}, \citenamefont {Kavoulakis},\ and\ \citenamefont {Reimann}}]{chergui2023superfluid}%
  \BibitemOpen
  \bibfield  {author} {\bibinfo {author} {\bibfnamefont {L.}~\bibnamefont {Chergui}}, \bibinfo {author} {\bibfnamefont {J.}~\bibnamefont {Bengtsson}}, \bibinfo {author} {\bibfnamefont {J.}~\bibnamefont {Bjerlin}}, \bibinfo {author} {\bibfnamefont {P.}~\bibnamefont {St{\"u}rmer}}, \bibinfo {author} {\bibfnamefont {G.}~\bibnamefont {Kavoulakis}}, \ and\ \bibinfo {author} {\bibfnamefont {S.~M.}\ \bibnamefont {Reimann}},\ }\href@noop {} {\bibfield  {journal} {\bibinfo  {journal} {arXiv preprint arXiv:2302.00385}\ } (\bibinfo {year} {2023})}\BibitemShut {NoStop}%
\end{thebibliography}%

\end{document}